\documentclass[12pt,a4paper]{article}
\usepackage[T1]{fontenc}
\usepackage[a4paper,margin=1in]{geometry}
\usepackage[margin=1in,font=footnotesize]{caption}
\usepackage{amsmath,amssymb,amsthm,bbm,url,color,natbib,tikz,enumerate,hyperref,pgfplots,titlesec}

\titleformat*{\section}{\large\bfseries}
\titleformat*{\subsection}{\large\bfseries}
\titleformat*{\paragraph}{\bfseries}

\usepgfplotslibrary{fillbetween}
\pgfplotsset{compat=1.18}

\newtheorem{lemma}{Lemma}
\newtheorem{theorem}{Theorem}

\newtheorem{proposition}{Proposition}

\newtheorem{assumption}{Assumption}

\renewcommand{\r}{r}
\renewcommand{\tt}{\varrho}

\newcommand{\topic}[1]{\textit{Content: } }

\newcommand{\pagetarget}[2]{\phantomsection\label{#1}\hypertarget{#1}{#2}}

\title{The Verification Bottleneck}
\author{Peter Achim\footnote{University of York, peter.achim@york.ac.uk}}
\date{October 31, 2026}
\begin{document}
\maketitle
\begin{abstract}
  This paper studies how generative AI, by lowering the cost of research production, affects scientific verification in a general-equilibrium model of peer review. Privately informed researchers choose observable output, an editor optimally assigns review depth, and reviewers supply time according to its expected scientific contribution. Costly output helps screen research quality, and cheaper production lets weaker researchers imitate stronger ones more readily. Beyond a technological threshold, reviewer capacity and verified knowledge weakly decline even under optimal editorial decisions, and strictly when a screening constraint has positive shadow value. Reviewer withdrawal can amplify the loss. With observable quality, the same progress weakly increases both.
\end{abstract}

\section{Introduction}
 
Scientific progress depends on both the production of new ideas and their independent verification. The stock of established knowledge grows through organized scepticism \citep{Merton1973}. Claims must withstand critical examination before they are accepted, and that examination requires scarce expert attention. Peer review, a central institution through which scientific communities organize the evaluation and certification of research, is under growing pressure. Major publishers report recent annual increases in submissions of around 20 percent or more, well above historical growth rates,\footnote{Elsevier reports article submissions rising by approximately 20\%, from 3.5 million in 2024 to more than 4.2 million in 2025 \citep{relx2025annual}, following substantial growth in the preceding year \citep{relx2024annual}. Springer Nature reports that submissions grew by more than 30\% to 3.1 million in 2025 \citep{springernature2025annual}. Annual submissions to arXiv increased by 14.3\%, from 248,823 in 2024 to 284,486 in 2025 \citep{arxiv_stats_2026}. By comparison, \citet{larsen2010} estimate that scientific publication volumes grew at approximately 4--6\% annually over much of the twentieth century.} while editors report growing difficulty recruiting qualified reviewers.\footnote{\citet{fox2017reviewers} report that, across six journals in ecology and evolution, the share of review invitations resulting in a completed review fell from approximately 56\% in 2003 to 37\% in 2015. \citet{kuenzli2022} document a similar decline at two public-health journals, and in a survey of Australian journal editors \citet{jamali2026peerreviewbreakingdown} find that recruiting suitable reviewers is widely regarded as a significant challenge. Industry data covering eight years of a major submission system show that review invitations more than doubled while the share accepted fell from 43\% in 2018 to 22\% in 2024 \citep{Silverchair2026}. Invitation acceptance rates measure responses to individual requests; aggregate reviewer capacity also depends on the size of the reviewer pool.} Concern has grown about whether scientific institutions can evaluate a rapidly expanding volume of research \citep{adam2025peer}.
 
With generative AI, these concerns take on a new dimension. AI increasingly performs cognitive tasks involved in research, which include in particular writing, programming, as well as reasoning, and data analysis. \citet{noy_zhang_2023} and \citet{brynjolfsson_li_raymond_2023} document substantial productivity gains in professional writing and other knowledge work, and \citet{novymarx_velikov_2025} demonstrate the potential for automating much of the production of research papers. In science, \citet{kusumegi_etal_2025} estimate increases in manuscript output of between 24 and 89 percent following LLM adoption, depending on field and researcher characteristics. In a survey of scientists reporting time savings from AI, 46 percent report spending more than a quarter of those savings checking AI outputs \citep{CodreanuEtAl2026}. Producing research and establishing its value are distinct economic activities, and they can change at different rates as technology improves. It is natural to ask how technological progress in research production affects the allocation and voluntary supply of scientific verification.
 
A key challenge is that generative AI can make it harder for journals to direct scarce reviewer attention towards the most valuable research, because the observable characteristics on which initial screening relies may be becoming less informative about underlying scientific quality. This is supported by recent evidence, which shows that among LLM-assisted manuscripts, more sophisticated writing is associated with lower subsequent publication success, reversing the relationship observed among human-written manuscripts \citep{kusumegi_etal_2025}. AI-generated papers can also appear plausible while the reasoning connecting their claims and evidence breaks down \citep{OhEtAl2026}. These developments raise the broader question how cheaper research production affect the screening of scientific work, the allocation of scarce scrutiny, and the willingness of reviewers to supply their time.

To study these forces,  I  develop a general-equilibrium model of peer review in which researchers, an editor, and reviewers jointly determine the production of verified knowledge. Researchers privately know their quality and choose the observable scope of their research. An editor commits to a rule that assigns review depth according to research scope. Review depth measures the intensity of scrutiny, while verification load measures the reviewer time required to carry it out. A broader project may require more reviewer time even when examined less deeply. Verified value is the expected scientific contribution established through review. Researchers value this contribution and bear the cost of developing their research. The editor allocates scrutiny to maximize aggregate verified value, subject to researchers' incentives and the available reviewer capacity. Reviewers voluntarily supply that capacity according to the expected scientific contribution of their time relative to its opportunity cost. The editorial rule is chosen optimally, so the analysis characterizes what can be achieved with the information and verification resources available.

 First, I show that asymmetric information distorts both research scope and the allocation of scrutiny, even when the editor chooses the optimal editorial policy. To discourage weaker researchers from imitating stronger ones, the editor assigns stronger research less scrutiny than its marginal scientific contribution would warrant, while weaker research receives more. The editor also makes stronger researchers undertake more scope-intensive assignments than would be efficient purely from the standpoint of reviewer time. These distortions arise because scrutiny serves two purposes: it establishes scientific knowledge and provides incentives for researchers to reveal their quality. Giving stronger research more scrutiny makes its assignment more attractive to weaker researchers, while giving weaker research more scrutiny helps keep them in their own assignments. Screening therefore diverts scarce reviewer attention from its most productive uses.

The second key finding is that distinguishing researchers becomes increasingly costly in reviewer time as research production becomes cheaper. Scope requirements help separate researchers because weaker researchers find it costly to imitate more ambitious work. Technological progress erodes this deterrent, making it necessary to impose increasingly demanding scope requirements to preserve a given difference in the benefits of editorial assignments. Because research scope must itself be verified, these requirements consume scarce reviewer time.  I  establish that maintaining a fixed incentive to imitate a stronger researcher's assignment, while continuing to establish positive scientific value from that assignment, eventually requires more reviewer time than any finite budget can provide. The editor must therefore reorganize scrutiny, reducing differences between assignments or pooling researchers whose quality can no longer be distinguished economically.

My main finding is that, under optimal editorial policies,  cheaper research production eventually causes both verified knowledge and the voluntary supply of reviewer time to cease increasing and potentially decline.  I  call this the verification bottleneck. As production becomes cheaper, incentive-compatible distinctions between researchers become harder to sustain, and the resulting screening distortions direct more scrutiny toward weaker research, reducing the average quality of work receiving reviewer attention. The editor can consequently establish less verified knowledge from a given stock of reviewer time, even after optimally adjusting research scope and scrutiny. Reviewers respond to the declining expected scientific contribution of their time by supplying less of it, which further reduces equilibrium verified knowledge. The decline can reach a plateau if the editor eventually pools researcher types because maintaining distinctions between them is no longer worthwhile.  I  also establish conditions under which differentiated assignments remain valuable and both reviewer capacity and verified knowledge continue to decline strictly indefinitely as production becomes cheaper.

 I  also examine the welfare implications of these results. Verified knowledge measures the scientific contribution established through peer review, while economic surplus additionally accounts for the real costs of producing research and supplying reviewer time. Cheaper production reduces these costs even when less knowledge is verified, so the effect on total surplus is ambiguous. Moreover, reviewers choose their participation according to the expected scientific contribution of their time, which need not equal its marginal social value net of production costs. Voluntary reviewer supply can consequently be above or below the level that maximizes surplus. The verification bottleneck therefore establishes a deterioration in the production of verified knowledge and the capacity available to produce it, without implying that every measure of welfare must decline.

Finally,  I  investigate how institutions can alleviate the verification bottleneck. Improvements in verification technology reduce the reviewer time needed to establish scientific value, while more informative preliminary screening allows editors to distinguish researchers without relying as heavily on costly scope requirements. Both changes increase attainable verified knowledge and support greater reviewer capacity. Scope-contingent submission fees provide another means of discouraging imitation; in my two-type illustration, they can restore the full-information allocation when production costs are sufficiently low.  I  also consider a competitive market in which researchers pay for the reviewer resources they use. In this benchmark, cheaper research production increases maximized surplus net of production and verification costs. These results identify the importance of verification technology, information, and institutional incentives in determining how scientific evaluation responds to advances in research production.

\paragraph{Related literature.}
Research on generative AI and scientific evaluation has grown rapidly since the release of ChatGPT in late 2022. Researchers have measured AI use in manuscripts and reviews, its association with editorial outcomes, and the quality of AI-generated assessments \citep{LiangEtAl2024,kusumegi_etal_2025,GartenbergEtAl2026,HuangKalaySchallheim2026,OhEtAl2026}. Theoretical work has largely examined the pressure created by submission volume, a concern that predates AI. \citet{Cotton2013} shows how submission fees and editorial delays screen authors by their prospects of acceptance and economize on refereeing resources. \citet{BergstromGross2026} model a feedback in which additional submissions strain review, reduce accuracy, and encourage further submissions, while \citet{LiuTan2026} examine how more precise review can interrupt it. \citet{Atasu2026} shows how AI-assisted submissions can congest reviewer capacity and degrade evaluation, and characterizes editorial responses involving fees, compensation, and review tracks. \citet{LopezLiraSeyfi2026} show that optimally chosen fees can prevent welfare losses from an AI-driven increase in submissions, and \citet{Lawford2026} and \citet{Fourie2026} study submission quotas and prices when manuscript production is cheap. 

Other work focuses on the review process itself: \citet{Hakobyan2026} shows how AI-generated reports can displace genuine reviewing effort, and \citet{Gans2026} studies how journals should share AI assessments with human reviewers.  I  study the information that submissions convey to the editor. Cheaper production makes scope a less effective screen of underlying quality, reducing the verified knowledge attainable from a given stock of reviewer time even under the editor's optimal response. The resulting decline in the scientific return to reviewing affects the supply of that scarce input. In this sense,  I  formalize for scientific certification the argument of \citet{CataliniHuiWu2026} that verification becomes the binding constraint as automation becomes cheaper. 
Relatedly, \citet{Lamba2026} derives verification time from sequential inference and studies its implications for the allocation of scarce expert attention through triage and organizational hierarchy.

The information conveyed by scope choices rests on costly signaling. Agents with different private characteristics can separate when the net returns to an observable action differ across types \citep{Spence1973Signaling}. In academic publishing, \citet{Ellison2002Standards} distinguishes the importance of a contribution from the craftsmanship of its presentation and explains how publication standards can place increasing weight on craftsmanship, a dimension that generative AI makes cheaper to produce. \citet{Condorelli2012} shows that when incentives must come from costly actions, the allocation itself becomes part of the incentive problem. \citet{GaldinSilbert2025} find that generative AI weakened the signalling value of tailored job applications and reduced employers' ability to identify stronger workers, and \citet{CetemenOzdenoren2026} show that separation can persist when AI generates scope, at the cost of a heavier signalling burden on higher types. In our setting the costly signal contributes directly to verified value, and evaluating it consumes scarce reviewer time. Cheaper signalling therefore changes both the allocation of scrutiny and the scientific contribution that scrutiny establishes.

The editor's problem also relates to mechanism design with costly verification. \citet{BenPorathDekelLipman2014CostlyVerif} study incentive-compatible allocations when a principal can verify agents' private information at a cost, and \citet{ErlansonKleiner2024} introduce a limit on the amount of verification, which corresponds most closely to our fixed reviewer capacity. Their models examine verification as a means of checking reports. In peer review verification also creates value: deeper scrutiny can establish a larger scientific contribution, while consuming more reviewer time. This productive role connects the analysis to certification. \citet{AlbanoLizzeri2001} study how certification affects incentives to supply quality, and \citet{BizzottoRudigerVigier2020} examine the interaction between applicants' disclosure and the certifier's acquisition of information. Closest to our approach, \citet{CelikStrausz2026} distinguish a certifier's screening of private information from its acquisition of new information about quality, and show how information rents lead a profit-maximizing certifier to restrict the latter. Our editor screens through assignments of scope and scrutiny, while scrutiny determines both the value established and the reviewer time required. Incentive constraints therefore distort the productive use of verification capacity.

Reviewer capacity is itself an equilibrium outcome in our model. \citet{EngersGans1998} develop an equilibrium account of voluntary refereeing in which reviewers care about journal quality, and show how that motivation affects the role of financial compensation. \citet{tite_schroter2007} find that a manuscript's expected contribution to its field is among the leading reported reasons for accepting a review invitation, and experiments by \citet{chetty_etal_2014} and \citet{squazzoni_etal_2013} show that reviewing responds to monetary and social incentives.  I  build on the scientific-value motive: reviewers supply time according to the expected verified value of their assignments, so the resources available to the editor depend on how effectively the editor uses the capacity already supplied. 

Our equilibrium mechanism also relates to general-equilibrium screening under adverse selection. Building on \citet{RothschildStiglitz1976}, \citet{GuerrieriShimerWright2010} study screening in competitive search, where matching probabilities and sorting depend on market conditions. \citet{AusterGottardi2024} show that greater meeting capacity changes the balance between sorting across markets and screening within them. Our model traces a comparative static from production technology through the allocation of scrutiny to the equilibrium supply of reviewer time. Cheaper production changes which assignments researchers are willing to select; those assignments determine the expected scientific return to reviewing; and that return determines the capacity available to the editor.
\section{Model}\label{sec:model}
This section develops the general model of idea production and verification, then microfounds the verification procedure as a dynamic experimentation problem, showing that the reduced-form structure used throughout is the canonical representation of a natural class of verification technologies.

\subsection{Environment}
There are three sets of agents. \emph{Researchers} produce and submit ideas. \emph{Reviewers} supply verification effort. An \emph{editor} allocates scarce verification capacity across submissions through an editorial mechanism that conditions on observable scope.

\paragraph{Researchers, submissions, and verification.}

A unit mass of researchers is partitioned into finitely many types $\theta_1<\cdots<\theta_M$, where type $\theta_i$ occurs with mass $f_i>0$ and $\sum_{i=1}^M f_i=1$. Types represent  all payoff-relevant private information the researcher has about the underlying quality or promise of a project affecting the submission and its evaluation. We denote the type space by $\Theta\equiv\{\theta_i\}_{i=1}^M$. Although only finitely many types occur, the primitives are defined on an open interval containing $\Theta$; derivatives with respect to $\theta$ refer to these smooth extensions.  After observing $\theta$, a researcher chooses $x \ge 0$, which I refer to as the scope of the submission. Scope is an observable, costly-to-produce characteristic of the attempted contribution that raises its potential value and affects its verification requirements.\footnote{Potential scientific value and verification requirements are distinct but correlated among submissions entering the refereeing process. Simple but high-value discoveries are scarce, while complicated, low-value work can often be identified and screened out at the editor’s desk.} We assume the editor observes scope at no cost.

  Before observing scope decisions, the editor commits to an editorial mechanism
\[
\tt=\bigl((x_i,r_i)\bigr)_{i=1}^M
\in
\mathbb R_+^M\times\mathbb R_+^M\equiv\cal M.
\]
Here $r_i$ is the review depth assigned to a submission of
realized scope $x_i$ from type $\theta_i$. 

\paragraph{Verified value and verification load.} Evaluating a submission has two distinct components we represent in reduced form. The first is that scrutiny converts a private claim into a public, certified piece of knowledge whose reliability has been established by independent assessment. Because the outcome of any individual test is uncertain ex ante, the relevant economic object is the expected social value generated when the submission must first pass a test of the prescribed depth. We call this object the \emph{verified value} of the evaluation, and write it as $v(x, r, \theta)\in\mathbb{R}_+$. Verified value depends on the submitter's type because higher types are more likely to produce claims that survive scrutiny, on scope because more ambitious research can offer greater potential scientific value, and on review depth because more thorough testing converts a larger fraction of correct claims into certified ones. 

We assume that verified value satisfies the following assumptions.

\begin{assumption}[Verified Value]\label{ass:v}
The verified-value function $v$ is twice differentiable, bounded, zero at zero scope and review depth, $
v(0, r, \theta)=v(x, 0, \theta)=0$, and satisfies
$v_x>0$, and $v_r>0$ for all $x,r>0$.
\end{assumption}

Assumption~\ref{ass:v} has two main implications. First, greater research scope increases expected scientific value because more ambitious research can yield findings with broader scientific implications. Second, the positive effect of scrutiny on verified value encodes academic scepticism. The underlying assumption is that for scientific contributions to be verified, the verification itself must produce evidence, which is more likely to be attained when research is evaluated at greater depth.

We occasionally consider a specific case 
in which verified value is multiplicatively separable
in researcher type and the research--review bundle. Formally, we say that a verified value function $v$ is \textit{separable by type} if it takes the form
\begin{equation}\label{eq:separable}
v(x,r,\theta)=a(\theta)y(x,r),
\end{equation}
where $a(\theta)>0$ is strictly increasing in type
and $y(x,r)$ is a common verification-success function. Under separability, researcher quality scales the value
of successful verification without changing how scope
and review depth jointly determine verification success.

The second component of the verification process is the amount of reviewer time the verification itself consumes.  We call the expected review time required for the verification the \emph{verification load}, and write it as $\ell(x, r, \theta)\in\mathbb{R}_+$. Verification load is the expected draw on the scarce input that reviewers supply. We assume that verification load satisfies the following assumptions. 
\begin{assumption}[Verification Load]\label{ass:ell}
The verification-load function $\ell$ is twice differentiable, zero at zero scope or review depth, $\ell(0, \r, \theta)=\ell(x, 0, \theta)=0$, and  $\ell_\r>0$ for all $x>0$ and $\r\ge0$. Moreover, for every $\varepsilon>0$,
\[
\lim_{K\to\infty}
\inf_{\substack{\theta\in\Theta,\;x,r\geq0\\
x+r\geq K,\;v(x, r, \theta)\geq\varepsilon}}
\ell(x, r, \theta)=\infty.
\]
\end{assumption}
Assumption~\ref{ass:ell} describes the resources consumed
by verification. The 
condition $\ell_r>0$ reflects the natural property that deeper scrutiny
requires more reviewer time. The final condition is a key coercivity assumption which says that verifying research of arbitrarily broad scope,
or subjecting a submission to arbitrarily intensive scrutiny,
requires unbounded reviewer time if the process is to produce
a meaningful amount of certified knowledge.  Essentially, it says that as either scope or review depth grows beyond bound, the verification load diverges. 

It is worthwhile to emphasize the distinction between review depth and verification load. Review depth is an input into the verification process which we assume is parameterized by a measure of review intensity. Verification
load, on the other hand, is a resource measure, representing the expected reviewer time needed to carry
out that review at given depth. The distinction will become transparent in the microfoundation of Subsection~\ref{subsec:experimentation}.

\paragraph{Technology and the cost of idea production.}

Producing a submission is costly. We represent the cost of producing a submission of scope $x$ for type $\theta$ by the drafting cost function $c(x,\theta;\tau)$, where $\tau$ indexes the state of production technology. Better technology lowers the cost of developing research scope. We assume that drafting costs satisfy the following assumptions. 
\begin{assumption}[Drafting cost]\label{ass:c}
The cost function is twice differentiable, equal to zero at zero scope, $c(0,\theta;\tau)=0$, and satisfies $c_x\ge0$ and $c_{x\tau}<0$ for all $x>0$. Moreover, marginal drafting costs vanish uniformly on bounded scope sets: for every finite $X$,
\[
\lim_{\tau\to\infty}\sup_{\theta\in\Theta,\;x\in[0,X]}|c_x(x,\theta;\tau)|=0.
\]
Finally, costs grow without bound as scope increases,
\[
\min_{\theta\in\Theta}c(x,\theta;\tau)\to\infty
\quad\text{as }x\to\infty.
\]

\end{assumption}
Assumption~\ref{ass:c} captures how production technology changes
the resources required to develop and elaborate research.
The condition $c_x\geq0$ reflects that developing greater
research scope requires additional effort or resources. The condition
$c_{x\tau}<0$ captures the productivity gains from technologies
that make these activities less costly at the margin.
Such gains are particularly relevant for tasks involved
in developing more ambitious research. Technological
progress therefore reduces the additional cost of increasing
research scope.

The remaining conditions describe the limits of technological
progress and research production. The assumption that marginal
drafting costs vanish as technology advances represents the
possibility that increasingly sophisticated tools can automate
an ever larger share of the activities involved in producing
research. In the limit, any given research scope can be
developed at negligible cost. 
The final condition requires costs to grow without bound
as scope increases at any fixed level of technology.
No given technology allows researchers to develop unbounded
research scope with finite effort and resources.

The researcher's private trade-off between the value of elaboration and its cost can be summarized by the net payoff
\[
u(x, \r, \theta;\tau):=v(x, \r, \theta)-c(x,\theta;\tau),
\]
which gives the expected return from research of scope $x$ when review depth is fixed at $r$. We impose the following assumption on the combined object $u$ directly. 

\begin{assumption}[Utility]\label{ass:u}
For every $\theta\in\Theta$, every $\tau$, and every
$r>0$, $\lim_{x\downarrow0}u_x>0$. Moreover, for all $x,r>0$,
\[
u_\theta>0,
\qquad
\frac{\partial}{\partial\theta}
\left(
\frac{u_x }
     {u_r }
\right)>0.
\]
\end{assumption}

The first condition in Assumption~\ref{ass:u} rules out a zero-scope
corner when a researcher receives positive review depth: starting
to develop a submission has positive marginal private value. It
does not require $u_x>0$ at every scope level; for a given finite
technology, drafting costs may eventually outweigh the marginal
benefit of further elaboration. The condition $u_\theta>0$ says that
higher types obtain higher private payoffs from the same positive
submission-review bundle.

The final condition is strict Spence--Mirrlees single crossing. Since
$u_r=v_r>0$, the ratio $u_x/u_r$ measures the marginal willingness
to give up review depth in exchange for greater scope while holding
private utility fixed. Higher types have a greater willingness to
make that trade. Incentive compatibility consequently orders
assigned scope weakly by type, although distinct types may still
receive the same bundle. This sorting property, together with the
technology effect on relative drafting costs, is what permits the
screening constraints to be organized into upward and downward
incentives.

\paragraph{Productivity of verification.}
The \emph{marginal product of verification} (MPV) measures the expected scientific value established by an additional unit of reviewer time, holding research scope fixed. For $x>0$, define
\begin{equation}\label{eq:MPV}
q(x,r,\theta):=\frac{v_r(x,r,\theta)}{\ell_r(x,r,\theta)}.
\end{equation}
The numerator measures the marginal scientific return to deeper scrutiny, while the denominator measures the reviewer time required to provide it. We impose two restrictions on how this marginal return varies with researcher quality and the allocation of scope and scrutiny.

\begin{assumption}[Verification Productivity]
\label{ass:q}
The marginal product of verification satisfies $q_\theta>0$. Moreover, for any type $\theta\in\Theta$ and any two positive feasible bundles $z=(x,r)$ and $z'=(x',r')$ satisfying $x<x'$ and $v(z,\theta)=v(z',\theta)$, we have
\[
\frac{v_r(z',\theta)}{\ell_r(z',\theta)}
\leq
\frac{v_x(z',\theta)}{\ell_x(z',\theta)}
\quad\Longrightarrow\quad
q(z,\theta)<q(z',\theta).
\]
\end{assumption}

Assumption~\ref{ass:q} orders marginal verification productivity across both researcher types and editorial assignments. First, at a given scope and review depth, higher-quality research generates more verified value per additional reviewer hour. Second, holding a researcher's verified value constant, greater scope and less scrutiny raise the marginal return to additional reviewing whenever the substitution toward scope locally saves reviewer time. The intuition is that increasing scope at the expense of scrutiny can initially raise verification productivity, but eventually the growing burden of verifying excessive scope may reverse this effect. Together, these properties ensure that the editor obtains a higher marginal return from allocating reviewer time to higher-quality research across the relevant scope--scrutiny assignments.

\paragraph{Aggregate outcomes.}

Given an editorial mechanism $\tt(\cdot)$ and the induced  choices $\{x_i\}_{i=1}^M$, aggregate verified value and aggregate load are
\[
V(\tt)
:=
\sum_{i=1}^M f_i\, v\bigl(x_i,r_i,\theta_i\bigr),
\qquad
L(\tt)
:=
\sum_{i=1}^M f_i\, \ell\bigl(x_i,r_i,\theta_i\bigr).
\]
We define \emph{impact} as the verified value per unit of reviewer capacity supplied.
For a menu $\tt$ implemented at reviewer capacity $T>0$, define
\[
I(\tt;T):=\frac{V(\tt)}{T}.
\]
A high impact value $I$ means that the available stock of reviewer capacity
produces substantial verified value, while a low value means that scarce
verification capacity produces comparatively little certified knowledge.
The editor need not exhaust capacity at every fixed-capacity problem;
unused capacity simply contributes no verified value and therefore lowers
impact.

Let 
\[
\bar V
:=
\sum_{i=1}^M f_i
\sup_{x,\r\geq0}
v(x, \r, \theta_i)
<\infty .
\]

\paragraph{Reviewers and verification supply.}

A unit mass of reviewers decides in aggregate how much total verification capacity $T \geq 0$ to make available.\footnote{The model treats reviewers as a separate population from researchers. In practice the same person often occupies both roles, but collapsing the two populations would reinforce the bottleneck, since a surge in submissions would pull agents toward production and away from review simultaneously.} The editor allocates submissions to reviewers according to the announced editorial mechanism $\tt(\cdot)$, and reviewers take the resulting pool composition as given when choosing how much capacity to supply. 

We postulate that the reviewers are motivated by the expected impact of the verification
capacity they supply, consistent with evidence that reviewing effort is
sustained primarily by professional norms and prosocial motivation
\citep{chetty_etal_2014,squazzoni_etal_2013}. A reviewer
decides whether to make time available before knowing which particular
submission will be assigned. Conditional on entering the reviewer pool,
assignments are drawn from the pool generated by the editorial
mechanism. The relevant return is therefore the expected verified value
generated per unit of reviewer capacity supplied,
\[
I=\frac{V}{T}.
\]

The payoff from supplying $T$ units of aggregate capacity is
\[
\Pi(T)=\phi I T-C(T;\tau),
\]
where $\phi>0$ is the weight placed on the expected contribution of a
review assignment and $C(T;\tau)$ is the aggregate opportunity cost of
supplying $T$ units of reviewer capacity.

\begin{assumption}[Review Cost]\label{ass:C}
The function $C(T;\tau)$ is twice continuously differentiable and satisfies
\[
C_T(T;\tau)>0 \quad \forall T>0,
\qquad
C_{TT}(T;\tau)>0,
\qquad
C_{T\tau}(T;\tau)=0,
\qquad
C_T(0;\tau)=0.
\]
Moreover,
\[
\lim_{T\to\infty}C_T(T;\tau)=\infty.
\]
\end{assumption}

Assumption~\ref{ass:C} describes the opportunity cost of
supplying reviewer time. Reviewing competes with research,
teaching, administrative responsibilities, and other
professional commitments. The assumption that marginal
costs are increasing reflects that additional reviewer
capacity becomes progressively more difficult to mobilize.
Researchers may be willing to accommodate occasional review
requests at relatively little cost, but supplying more
reviewer time requires increasingly substantial sacrifices
of alternative activities. The condition $C_T(0)=0$
captures the negligible marginal cost of supplying a very
small amount of review, while the requirement that marginal
costs eventually grow without bound reflects the finite
time and attention available to the research community.

The condition $C_{T\tau}=0$ reflects an asymmetry between
research production and verification. Technologies that
reduce the cost of writing, coding, and elaborating research
need not reduce the opportunity cost of the independent
scrutiny required to certify it. In particular, the ability
to produce a manuscript more quickly does not necessarily
make the time devoted to reviewing that manuscript less
costly. The assumption isolates a setting in which
production technology improves while the marginal cost
of supplying human reviewer attention remains unchanged.
Reviewer capacity nevertheless responds endogenously to
technological progress through changes in the expected
scientific return to reviewing.

Reviewers choose capacity before knowing the particular submissions
they will evaluate. Their marginal incentive to supply time is
therefore determined by expected \emph{average} verified value per
unit of supplied capacity, rather than the derivative of total
verified value with respect to capacity. Convex reviewer costs do
not, by themselves, make equilibrium unique, which is why the
analysis selects the equilibrium with greatest reviewer capacity.
If reviewers could condition their supply or assignment decisions on
observable manuscript characteristics, an additional allocation
problem would arise on the reviewer side; the baseline isolates the
aggregate capacity response.

\paragraph{Equilibrium concept.} The editor commits to a menu and allocates the available reviewer
capacity across submissions. Reviewers choose how much capacity to
supply given the impact they expect their effort to generate. An
equilibrium is a configuration in which the editor's menu is optimal
given the prevailing reviewer capacity, reviewer supply is optimal
given the expected impact, and the impact taken as given by reviewers
is the impact actually generated by the editor's menu.

Formally, a pure-strategy equilibrium consists of an editorial mechanism \(\varrho^* = ((x_i^*, \r_i^*))_{i=1}^M\), reviewer capacity
\(T^* > 0\), and impact \(I^*\) such that:
\begin{enumerate}[(i)]
\item \emph{Researcher optimality.} For every \(i,j\), \[u(x_i^*, r_i^*, \theta_i;\tau)\ge u(x_j^*, r_j^*, \theta_i;\tau),\] and \[u(x_i^*, r_i^*, \theta_i;\tau)\ge0.\]
\item \emph{Editorial optimality.} Given \(T^*\), the menu \(\varrho^*\) maximizes \(V(\varrho)\) subject to \(L(\varrho)\le T^*\), incentive compatibility, and participation.
\item \emph{Reviewer optimality.}
\[
T^* \in \arg\max_{T \ge 0}\,\{\phi I^* T - C(T; \tau)\}.
\]
\item \emph{Consistent impact.}
\[
I^* = \frac{V(\varrho^*)}{T^*}.
\]
\end{enumerate}

Conditions (i) and (ii) describe the editor's problem. The menu is
incentive compatible and optimally allocates the reviewer capacity
\(T^*\) treated as fixed. Condition (iii) describes the reviewer
problem. Capacity \(T^*\) is the supply response to the expected
impact \(I^*\). Condition (iv) closes the loop by requiring that the
impact reviewers anticipate is the impact the editor's menu actually
delivers. Because the menu is direct, researcher optimality is encoded in incentive compatibility together with participation, so we do not need a separate best-response condition for researchers' submission choices.
\subsection{Microfoundation: exponential verification}
\label{subsec:experimentation}

The reduced-form notions of verified value and verification load arise naturally
from standard models of dynamic evidence production. Many classes of
information-generating processes, including Wald-type sequential testing and
more general experimentation models, give rise to these objects: a reviewer
dynamically acquires evidence about a submission until enough information has
accumulated to accept or reject it. The expected value of the outcome of that
test is what we refer to as verified value, while the expected reviewer time
consumed by the test under the induced stopping rule is the verification load.
We illustrate this using a single-agent variant of the exponential bandit of \citet{KellerRadyCripps2005}, as
it yields simple closed-form expressions and a familiar dynamic structure. 

Suppose a researcher of type $\theta$ produces a good idea with probability $\pi(x,\theta)$, where $x$ is observable research scope. Conditional on a good idea, decisive evidence arrives at rate $\lambda/x$ per unit of reviewer time, with $\lambda>0$. Conditional on a bad idea, decisive validating evidence never arrives. The editor prescribes review depth $r$, corresponding to a planned reviewer-time budget $xr$. The probability of confirming a good idea is therefore $1-e^{-\lambda r}$, giving expected verified value
\begin{equation}\label{eq:v_exp}
v(x, r, \theta)=\pi(x,\theta)(1-e^{-\lambda r}).
\end{equation}

Finding decisive evidence can eliminate some verification work, although certain checks may still be necessary. Let $\alpha\geq0$ govern how much work remains after discovery. For work scheduled $t$ units of review depth before the prescribed horizon, the fraction that can be avoided following discovery is $e^{-\alpha\lambda t}$. Conditional on a sound idea, the probability that decisive evidence has arrived before this work is reached is $1-e^{-\lambda(r-t)}$. Integrating the expected savings over the review horizon gives
\[
h_\alpha(r)=\int_0^r(1-e^{-\lambda(r-t)})e^{-\alpha\lambda t}\,dt.
\]
Expected verification load is
\begin{equation}\label{eq:ell_exp}
\ell_\alpha(x, r, \theta)=x\left[r-\pi(x,\theta)h_\alpha(r)\right].
\end{equation}

The parameter $\alpha$ measures how quickly the amount of avoidable work declines with its distance from the review horizon. At $\alpha=0$, all remaining work can be avoided following discovery, recovering the original exponential-bandit stopping rule of \citet{KellerRadyCripps2005}. Higher values of $\alpha$ leave more verification work to be completed after decisive evidence arrives. This changes the relative reviewer-time efficiency of scope and review depth.

\begin{proposition}[Consistency of the verification technology]
\label{prop:microfoundation}
Suppose $\pi(x,\theta)=a(\theta)p(x)$, where $a(\theta)\in(0,1)$ is twice continuously differentiable with $a'>0$, and $p:[0,\infty)\to[0,1)$ is twice continuously differentiable and satisfies
\[
p(0)=0,\qquad p'>0,\qquad -xp''\geq pp'\quad(x>0).
\]
Let $c(x,\theta;\tau)=e^{-\tau}\kappa(x)$, where $\kappa$ is twice continuously differentiable, $\kappa(0)=\kappa'(0)=0$, $\kappa'>0$ for $x>0$, and $\kappa(x)\to\infty$ as $x\to\infty$, and let $C(T)=T^2/2$. Then for every $\alpha\geq0$, the verification technology \eqref{eq:v_exp}--\eqref{eq:ell_exp} satisfies Assumptions~\ref{ass:v}--\ref{ass:C}. 
\end{proposition}

\hyperlink{proof_prop:microfoundation}{The proof is in Appendix~\ref{sec:proofs} on page \pageref{proof_prop:microfoundation}}. The proposition establishes the joint consistency of the assumptions for a class of stochastic verification technologies. The class permits flexible scope-productivity functions, including $p(x)=x/(b+x)$ and $p(x)=1-e^{-kx}$ for $b,k>0$. The parameter $\alpha$ governs the comparative advantage in verification formats and determines whether the additional conditions for persistent strict deterioration hold.

\section{Allocation of Scrutiny}
\label{sec:scarcity}

The editor allocates scarce reviewer attention to research of privately known quality. Scope and review depth determine the verified knowledge established by scrutiny; together, they also determine which assignments researchers are willing to choose. We begin with the editor's allocation problem, use observed quality to characterize efficient verification, and then show how screening changes the allocation.

\subsection{Editor's problem}

Fix reviewer capacity $T$ and production technology $\tau$, and let $z_i=(x_i,r_i)$ denote the scope and review depth assigned to type $\theta_i$. The editor chooses a menu $\varrho=(z_i)_{i=1}^M$ to maximize verified knowledge:
\begin{equation}\label{eq:editor-problem}
\begin{aligned}
 V^*(T,\tau):=\max_{\varrho}\quad
 &\sum_{i=1}^M f_i v(z_i,\theta_i)\\
 \text{subject to}\quad
 &u(z_i,\theta_i;\tau)\geq u(z_j,\theta_i;\tau)
 &&\text{for all }i,j,\quad (\lambda_{ij})\\
 &u(z_i,\theta_i;\tau)\geq 0
 &&\text{for all }i,\quad (\mu_i)\\
 &\sum_{i=1}^M f_i\ell(z_i,\theta_i)\leq T
 &&\hspace{2.8em}(\gamma).
\end{aligned}
\end{equation}
The first constraints make each type prefer her designated assignment to the alternatives. Participation requires that the value of the assigned scrutiny cover drafting costs. The capacity constraint limits the total reviewer time consumed. The quantities in parentheses are the multipliers on these constraints. The capacity multiplier $\gamma$ is the shadow value of an additional unit of reviewer time, equal to $V_T^*(T,\tau)$ where the value function is differentiable. The drafting technology $\tau$ changes the incentive and participation constraints through $c(x,\theta;\tau)$; the functions $v$ and $\ell$ describe the knowledge and reviewer time associated with a given bundle.

\subsection{Efficient verification}

Suppose the editor observes researcher quality. She can assign each type its appropriate research--review bundle, and the editor's problem~\eqref{eq:editor-problem} then has the capacity and participation constraints alone. At an interior optimum with slack participation, differentiating its Lagrangian with respect to the review depth and scope of type $i$ gives
\[
 f_i(v_r-\gamma\ell_r)(z_i,\theta_i)=0,
 \qquad
 f_i(v_x-\gamma\ell_x)(z_i,\theta_i)=0.
\]
The first condition yields, using the marginal product of verification in~\eqref{eq:MPV},
\[
 q(z_i,\theta_i)=\frac{v_r(z_i,\theta_i)}{\ell_r(z_i,\theta_i)}=\gamma.
\]
The marginal product of reviewer time is thus equalized across the submissions receiving positive scrutiny. An hour assigned to any one of them yields the same marginal verified value as an hour assigned elsewhere.

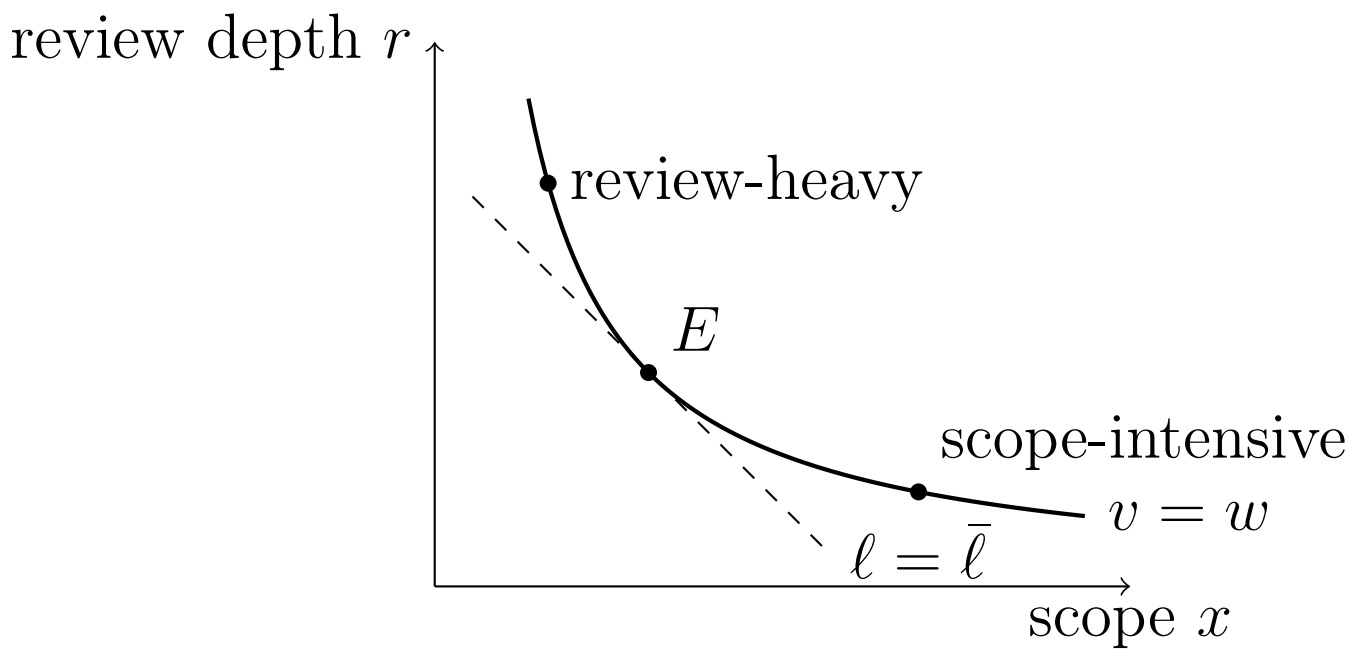
\begin{figure}[t]
\centering
\begin{tikzpicture}[scale=1.0]
\draw[->] (0,0) -- (4.6,0) node[below] {scope $x$};
\draw[->] (0,0) -- (0,3.6) node[left] {review depth $r$};
\draw[thick,domain=0.62:4.3,samples=100] plot (\x,{2/\x}) node[right] {$v=w$};
\draw[dashed,domain=0.25:2.6] plot (\x,{2.8284-\x}) node[right] {$\ell=\bar\ell$};
\fill (1.4142,1.4142) circle (1.6pt) node[above right] {$E$};
\fill (3.2,0.625) circle (1.6pt) node[above right] {scope-intensive};
\fill (0.75,2.6667) circle (1.6pt) node[right] {review-heavy};
\end{tikzpicture}
\caption{The efficient format. Bundles on the solid curve deliver the same verified value; the dashed curve is tangent to it at $E$, the bundle that achieves that value with least reviewer time. Scope-intensive and review-heavy formats lie on opposite sides of $E$.}
\label{fig:format}
\end{figure}

The second condition determines the research--review format. Holding type $i$'s verified value fixed, a marginal increase in scope permits a reduction in review depth of $dr/dx=-v_x/v_r$. The resulting change in reviewer load is $\ell_x-\ell_r v_x/v_r$. Combining the two first-order conditions gives
\begin{equation}\label{eq:efficient-format}
 \ell_x(z_i,\theta_i)
 -\frac{v_x(z_i,\theta_i)}{v_r(z_i,\theta_i)}
   \ell_r(z_i,\theta_i)=0.
\end{equation}
The editor therefore chooses, for each assigned level of verified value, the mix of scope and scrutiny that uses the least reviewer time. Figure~\ref{fig:format} depicts this \emph{efficient format} as the tangency between an iso-value curve and an iso-load curve. The editor chooses both how much knowledge to verify for each type and the format in which to verify it.

Write $V^{FI}(T,\tau)$ for the value of~\eqref{eq:editor-problem} when the editor observes quality, retaining participation. Write $V^{FB}(T)$ for the further benchmark in which participation is slack by assumption. As drafting costs fall, the former approaches the latter, whose allocation depends only on the verification technology and reviewer capacity. These benchmarks will identify the contribution of private information to the technological comparative static in Section~\ref{sec:reorganization}.

\subsection{Screening distortions}

Under private information, the editor's assignments must satisfy the incentive constraints in~\eqref{eq:editor-problem}. Stronger researchers are more willing to substitute additional scope for review depth. Scope requirements therefore help direct scrutiny toward stronger work: a demanding bundle is comparatively attractive to the researcher for whom that scope is less costly. The same screening motive affects the review attached to each bundle. Deeper scrutiny increases the private value of an assignment, including its attraction to researchers for whom it was intended as an alternative.

To see how these incentives alter the allocation, consider two types $\theta_L<\theta_H$ receiving distinct positive bundles. Suppose the weak researcher's constraint against selecting the strong bundle,
\begin{equation}\label{eq:screening-upward-ic}
 u(z_L,\theta_L;\tau)\geq u(z_H,\theta_L;\tau),
\end{equation}
binds with multiplier $\lambda>0$, while participation and the strong researcher's downward constraint are slack. These are the constraints governing distinct bundles at sufficiently advanced technology. In this case the Lagrangian of the editor's problem~\eqref{eq:editor-problem} is
\[
 \mathcal{J}=\sum_{i=L,H}f_i v(z_i,\theta_i)
 +\lambda\bigl[u(z_L,\theta_L;\tau)-u(z_H,\theta_L;\tau)\bigr]
 +\gamma\left[T-\sum_{i=L,H}f_i\ell(z_i,\theta_i)\right].
\]
Differentiating with respect to $r_H$ and $r_L$, and using $u_r=v_r$, gives
\begin{equation}\label{eq:screening-review-foc}
\begin{aligned}
 f_H\bigl[v_r(z_H,\theta_H)-\gamma\ell_r(z_H,\theta_H)\bigr]
 &=\lambda v_r(z_H,\theta_L)>0,\\
 f_L\bigl[v_r(z_L,\theta_L)-\gamma\ell_r(z_L,\theta_L)\bigr]
 &=-\lambda v_r(z_L,\theta_L)<0.
\end{aligned}
\end{equation}
The two signs express opposite incentive effects of scrutiny. More review of the strong bundle makes imitation more rewarding for the weak researcher; the editor consequently holds back scrutiny of strong research even while its marginal product exceeds the value of reviewer time, $q(z_H,\theta_H)>\gamma$. More review of the weak bundle makes the weak researcher more willing to remain with her assignment; the editor consequently continues to review weak research even after its marginal product falls below $\gamma$. Thus~\eqref{eq:screening-review-foc} implies
\[
 q(z_H,\theta_H)>\gamma>q(z_L,\theta_L).
\]
The strong bundle is \emph{under-scrutinized}, and the weak bundle is \emph{over-scrutinized}, relative to the full-information condition $q=\gamma$.

The scope requirement on strong research provides another way to deter imitation. As in the efficient-format condition~\eqref{eq:efficient-format}, increasing scope while holding the strong researcher's verified value fixed permits a reduction in review depth at the rate $dr_H/dx_H=-v_x(z_H,\theta_H)/v_r(z_H,\theta_H)$. Differentiating the Lagrangian of~\eqref{eq:editor-problem} with respect to $x_H$ and subtracting $v_x(z_H,\theta_H)/v_r(z_H,\theta_H)$ times the first review-depth condition in~\eqref{eq:screening-review-foc} gives
\begin{equation}\label{eq:screening-scope-intensive}
\begin{aligned}
&f_H\gamma\left[
\ell_x(z_H,\theta_H)
-\frac{v_x(z_H,\theta_H)}{v_r(z_H,\theta_H)}\ell_r(z_H,\theta_H)
\right]\\
&\qquad=\lambda\left[
\frac{v_x(z_H,\theta_H)}{v_r(z_H,\theta_H)}u_r(z_H,\theta_L;\tau)
-u_x(z_H,\theta_L;\tau)
\right]>0.
\end{aligned}
\end{equation}
The inequality follows from single crossing: the weak researcher values additional scope relative to the review forgone less than the strong researcher does. Substituting scope for scrutiny consequently makes the strong bundle less attractive to the weak researcher. The expression on the left measures the associated change in reviewer load along the strong researcher's iso-verified-value curve. It is zero at the efficient format and positive under screening. The editor assigns an \emph{scope-intensive} bundle because the reviewer-time cost of additional scope is worth bearing to deter imitation.

Under separable verified value~\eqref{eq:separable}, the same combination of the weak bundle's scope and review-depth conditions yields
\begin{equation}\label{eq:screening-review-heavy}
\begin{aligned}
&f_L\gamma\left[
\ell_x(z_L,\theta_L)
-\frac{v_x(z_L,\theta_L)}{v_r(z_L,\theta_L)}\ell_r(z_L,\theta_L)
\right]\\
&\qquad=\lambda\left[
 u_x(z_L,\theta_L;\tau)
-\frac{v_x(z_L,\theta_L)}{v_r(z_L,\theta_L)}u_r(z_L,\theta_L;\tau)
\right]
=-\lambda c_x(x_L,\theta_L;\tau)\leq0.
\end{aligned}
\end{equation}
At fixed verified value, substituting scope for scrutiny would save reviewer time on the weak bundle. It would also reduce the weak researcher's utility by raising her drafting cost, making the strong bundle more attractive. The editor instead chooses a \emph{review-heavy} assignment to help keep the weak researcher in her own bundle.

The two-type first-order conditions reveal the mechanism. For any finite number of types, incentive compatibility orders bundles by scope and allows consecutive types to pool. The general proof sums the first-order conditions over the types sharing each bundle. The highest block faces imitation from below, and the lowest block consists of researchers tempted by higher assignments, so the same directional distortions follow.

The result concerns the reviewer capacities relevant for the subsequent equilibrium analysis. Lemma~\ref{lem:uniform-optima} establishes that, beyond a finite technology threshold, all equilibrium capacities lie in some interval $[T_-,T_+]$ with $0<T_-<T_+<\infty$. These bounds also give uniform interiority and slack participation of optimal assignments; the appendix establishes the required first-order conditions and the eventual slackness of downward constraints. Thus a single technological threshold works throughout the range of capacities that reviewers can supply in equilibrium.

\begin{proposition}[Screening distortions]
\label{prop:screening-distortions}
Fix any reviewer capacity $T>0$. Under Assumptions~\ref{ass:v}--\ref{ass:q}, at sufficiently advanced production technology,
the strongest researchers receive a scope-intensive,
under-scrutinized assignment, while the weakest
researchers receive an over-scrutinized assignment. 
Under separable verified value \eqref{eq:separable}, the weakest
researchers' assignment is also review-heavy.
\end{proposition}

\hyperlink{proof_prop:screening-distortions}{The proof is in the appendix  on page \pageref{proof_prop:screening-distortions}}. The directional inequalities become strict when the corresponding incentive constraints have positive multipliers; the proposition states weak distortions to cover pooling and slack screening constraints as well.

The direction of these distortions contrasts with standard screening models with monetary transfers, such as \citet{MussaRosen1978Quality}. There, the principal uses transfers to provide information rents, leaving the highest type's allocation efficient while distorting the allocations of lower types. Here, the editor provides incentives through scope requirements and review depth, both of which affect the production of verified knowledge and the use of scarce reviewer time. Screening therefore distorts the allocation at both ends. Strong researchers receive too little scrutiny relative to its marginal productivity, and their assignments require more scope relative to review depth than is reviewer-efficient. Weak researchers receive too much scrutiny, and under separability their assignments are tilted toward review rather than scope. These distortions arise because deeper scrutiny of strong research encourages imitation, whereas deeper scrutiny of weak research discourages it. The editor consequently reallocates verification away from where it produces the most knowledge and toward where it sustains incentive compatibility.

\section{Technology and the Limits of Screening}
\label{sec:reorganization}

Section~\ref{sec:scarcity} established how the editor uses scope and scrutiny to separate privately informed researchers. Production technology changes the cost of that separation. A lower drafting cost makes a given scope requirement easier to meet, and thereby changes which research--review assignments researchers are willing to choose. We study this effect on the editor's problem~\eqref{eq:editor-problem}, holding reviewer capacity fixed. The argument moves from the cost of sustaining a scope distinction to the editor's optimal response, and then, under separable verified value, to the allocation that emerges as drafting costs vanish.

\subsection{Cost of Separation}
\label{subsec:technology-incentives}

Suppose a weaker type $i$ receives $z_i=(x_i,r_i)$ and a stronger type $j$ receives $z_j=(x_j,r_j)$, with $x_i<x_j$. The weak type's incentive constraint in~\eqref{eq:editor-problem}, after substituting $u=v-c$, is
\begin{equation}\label{eq:upwardIC}
 v(z_j,\theta_i)-v(z_i,\theta_i)
 \leq c(x_j,\theta_i;\tau)-c(x_i,\theta_i;\tau).
\end{equation}
The left side is the weak researcher's gross gain from receiving the scrutiny attached to the strong bundle. The right side is the extra production cost of obtaining that bundle. It is this difference in drafting costs that allows the editor to direct more valuable scrutiny to the stronger researcher.

As $\tau$ rises, the cost difference on the right shrinks at any fixed pair of scope choices. To preserve a given advantage in verified value for the strong bundle, the editor must make it harder to imitate. She can require more scope from the strong researcher, or lower the review attached to the strong bundle and raise the review attached to the weak bundle. The first adjustment restores a private deterrent; the others erode the distinction in verification that the editor was trying to make. Since every scope requirement also changes the material that reviewers must examine, the private deterrent has a cost in reviewer time.

The technological restriction is particularly sharp if the editor tries to preserve a fixed gross gain from imitation. Write $L(\varrho)$ for the reviewer load of a menu, as in Section~\ref{sec:model}.

\begin{theorem}[Reviewer cost of separation]
\label{thm:cost-dependent-separation}
Under Assumptions~\ref{ass:v}--\ref{ass:c}, fix $i<j$ and $\delta,\varepsilon>0$.
For every finite reviewer budget $B>0$, there exists a finite $\tau_B$ such that, for all $\tau>\tau_B$, every incentive-compatible menu $\rho=(z_i)_{i=1}^M$ satisfying
\[
 v(z_j,\theta_i)-v(z_i,\theta_i)\geq\delta,
 \qquad v(z_j,\theta_j)\geq\varepsilon
\]
requires reviewer load $L(\varrho)>B$.
\end{theorem}

\hyperlink{proof_thm:cost-dependent-separation}{The proof is in the appendix on page \pageref{proof_thm:cost-dependent-separation}.} To see the force of the result, suppose the editor could maintain these two inequalities with a bounded reviewer budget. Positive verified value for type $j$, together with verification-load coercivity in Assumption~\ref{ass:ell}, would keep her scope and review depth bounded. But marginal drafting costs vanish uniformly on bounded scope sets. The right side of~\eqref{eq:upwardIC} would therefore converge to zero, contradicting the fixed gain $\delta$ on its left. The proof in Appendix~\ref{sec:proofs} makes this argument for any incentive-compatible menu.

The theorem identifies the resource cost behind the breakdown of scope-based screening. Strong researchers can continue to signal their type by undertaking larger scope requirements, but the editor must devote reviewer time to processing that scope if it is to generate positive verified knowledge. At any fixed reviewer budget, the editor therefore has to reduce the reward for obtaining a stronger bundle or reorganize which bundles she offers. A stronger statement for fixed allocations of verified value appears in Appendix~\ref{sec:proofs}.

\subsection{Productivity of Reviewer Time}
\label{subsec:fixed-reviewer-productivity}

Theorem~\ref{thm:cost-dependent-separation} fixes the weak researcher's gross gain from imitating a stronger bundle. In the editor's actual problem, that gain is endogenous. She can narrow the gap in verification between types until the imitation gain itself approaches zero. Such menus can require bounded reviewer time even as a fixed gain becomes prohibitively expensive. She can also pool types or redistribute scrutiny across assignments. The theorem therefore raises a further question: how much verified knowledge can the editor attain after choosing the best response to cheaper drafting?

The answer depends on which constraints bind at the optimum of~\eqref{eq:editor-problem}. Cheaper drafting relaxes researchers' participation and changes their incentives to imitate bundles with more or less scope. We need to establish which of those opposing effects remains relevant after the editor has reorganized the menu.

The source of the comparative static is visible in the editor's constraints. Technology enters~\eqref{eq:editor-problem} through drafting costs in participation and incentive compatibility. Differentiating these constraints with respect to $\tau$ and applying the envelope theorem at a regular, differentiable optimum gives
\begin{equation}\label{eq:envelope}
 V_\tau^*(T,\tau)
 =-\sum_{i=1}^M\mu_i c_\tau(x_i,\theta_i;\tau)
 +\sum_{i,j}\lambda_{ij}
   \int_{x_i}^{x_j}c_{x\tau}(s,\theta_i;\tau)\,ds.
\end{equation}
The first term is nonnegative: cheaper scope relaxes participation. The incentive terms depend on the direction of imitation. For a stronger researcher considering a lower-scope assignment ($x_j<x_i$), the integral is positive; lower drafting costs make her own high-scope assignment more attractive. For a weaker researcher tempted by a higher-scope assignment ($x_j>x_i$), the integral is negative. The drafting-cost advantage that kept her in her own bundle has diminished. This is the same deterrence effect isolated by~\eqref{eq:upwardIC}.

The decisive point is the direction of the remaining incentive constraints. At sufficiently advanced technology, participation becomes slack: every researcher obtains positive verified value, while the cost of producing a bounded assigned bundle approaches zero. The same high-technology region rules out binding downward constraints between distinct bundles. To see why, consider two separated types. If the strong researcher were just indifferent between her own bundle and the weak researcher's, the review-depth conditions derived from~\eqref{eq:editor-problem} would require the marginal product of review on the weak bundle to exceed that on the strong bundle: $q_L\geq\gamma\geq q_H$, with strict inequalities if that constraint carries a positive multiplier. At negligible drafting cost, however, the bundles are approximately on the strong type's iso-verified-value curve. Editorial efficiency along that curve and comparative verification advantage, Assumption~\ref{ass:q}, imply $q_L<q_H$. The resulting contradiction rules out a costly downward constraint. Lemma~\ref{lem:downward} extends the argument to chains of constraints and pooling blocks for arbitrary $M$.

The surviving cross-bundle constraints run upward: weaker researchers prefer their own assignment to the more scope-intensive bundles intended for stronger ones. For any fixed menu, each such constraint becomes harder to satisfy as $\tau$ increases, by~\eqref{eq:upwardIC}. This has a direct implication for the editor's \emph{optimal value}. Take a menu optimal at a sufficiently advanced technology $\tau_2$, and consider a slightly earlier technology $\tau_1<\tau_2$. Its binding upward constraints are relaxed at $\tau_1$; participation and downward constraints remain slack; and pooled types still receive identical bundles. Thus the editor can implement the later menu at the earlier technology, using exactly the same reviewer time and producing exactly the same verified value. The editor's optimum at $\tau_1$ must be at least as high. 

\begin{theorem}[Declining productivity of reviewer time]
\label{thm:fixed-capacity}
Under Assumptions~\ref{ass:v}--\ref{ass:C}, there is a finite $\bar\tau$ such that
\[
 V^*(T,\tau_2)\leq V^*(T,\tau_1)
 \qquad\text{for all }T\in\mathcal T
 \text{ and }\tau_2>\tau_1>\bar\tau.
\]
\end{theorem}

\hyperlink{proof_thm:fixed-capacity}{The proof is in the appendix on page \pageref{proof_thm:fixed-capacity}.} The theorem turns the rising cost of maintaining separation into a loss of \emph{institutional productivity}. The editor can respond by narrowing verification differences or pooling some researchers; these adjustments conserve reviewer time but sacrifice information that previously helped her direct scrutiny to valuable work. Even when she chooses the best such adjustment, a fixed stock of reviewer time eventually establishes weakly less knowledge as production becomes cheaper. The result concerns the editor's optimal use of the same reviewer capacity, before any withdrawal by reviewers.

The economic cost arises from directing reviewer time according to imitation incentives rather than marginal verification productivity, as in Section~\ref{sec:scarcity}. The uniform comparison in Theorem~\ref{thm:fixed-capacity} is what permits the equilibrium analysis of reviewer supply in Section~\ref{sec:bottleneck}. Conditions for strict deterioration are developed there.

Under full information the comparison runs in the opposite direction. Observing types removes the incentive constraints from~\eqref{eq:editor-problem}; technological progress then relaxes participation and weakly raises $V^{FI}(T,\tau)$. Eventually the participation constraints also cease to bind, and the full-information value equals $V^{FB}(T)$, which depends only on verification technology and capacity (Lemma~\ref{lem:fi-frontier}). At fixed reviewer capacity, the gap between the full-information and privately informed editor's attainable verified knowledge therefore grows at sufficiently advanced production technologies. The widening gap measures how much more valuable information about researcher quality becomes when scope loses its screening power.

The inequality becomes an equality when the editor has pooled types or when remaining differentiated assignments are efficient without a binding cross-type screening cost. Section~\ref{subsec:strict} gives a primitive sufficient condition for strict equilibrium deterioration. The next two results characterize how the editor reorganizes assignments when verified value is separable.

\subsection{Verification under Separability}
\label{subsec:separable-value}
\label{subsec:pooling-differentiation}

Theorem~\ref{thm:fixed-capacity} establishes that
cheaper production eventually reduces the verified
knowledge attainable from a fixed stock of reviewer
time. We now ask how the editor adjusts the allocation
across researchers as the ability to screen them
through scope requirements erodes. We focus on the case of separable verified value~\eqref{eq:separable}, 
\[
v(x_i,r_i,\theta_i)=a(\theta_i)y(x_i,r_i). 
\]
Separability by type allows us
to distinguish the verification success delivered
by an assignment from the scientific value produced
when verification succeeds. 

Consider two types $i<j$, with assignments
$(x_i,r_i)$ and $(x_j,r_j)$, and write
$y_i=y(x_i,r_i)$ and $y_j=y(x_j,r_j)$.
Incentive compatibility requires each researcher
to prefer her own assignment. Combining these
two inequalities gives
\begin{equation}\label{eq:ci-ic-interval}
\frac{c(x_j,\theta_j;\tau)-c(x_i,\theta_j;\tau)}{a_j}
\leq y_j-y_i \leq
\frac{c(x_j,\theta_i;\tau)-c(x_i,\theta_i;\tau)}{a_i}.
\end{equation}

The inequalities show how drafting costs sustain
differences in verification success. To offer
stronger researchers higher success, the editor
must require enough additional scope to deter
weaker researchers from imitating them. Yet at
bounded scope requirements, the additional cost
of producing that scope vanishes as technology
improves. Both bounds in~\eqref{eq:ci-ic-interval}
therefore converge to zero.

Optimal assignments remain bounded at fixed
reviewer capacity. Consequently, incentive
compatibility forces the verification-success
indices of all researcher types to converge.
The common level is determined by the editor's
limiting verified knowledge: writing
$A=\sum_i f_i a_i$, aggregate verified knowledge
is $\sum_i f_i a_i y_i$, so convergence of the
success indices implies that their common limit
is $y_T=\lim_{\tau\to\infty}V^*(T,\tau)/A$.

\begin{theorem}[Equalization of verification success]
\label{thm:ci-equalization}
Under Assumptions~\ref{ass:v}--\ref{ass:C} and separable verified value, fix $T\in\mathcal T$. As production costs vanish, every optimal editorial allocation satisfies
\[
 y\bigl(z_i^*(T,\tau)\bigr)\longrightarrow y_T
 \qquad\text{for every }i,
\]
where
\[
 y_T=\frac{\lim_{\tau\to\infty}V^*(T,\tau)}{\sum_i f_i a_i}.
\]
\end{theorem}

\hyperlink{proof_thm:ci-equalization}{The proof is in the appendix on page \pageref{proof_thm:ci-equalization}.}  
Theorem~\ref{thm:ci-equalization} establishes that
differences in verification success vanish as production
becomes cheaper. This does not imply that the editor
pools researcher types. The incentive constraints
in~\eqref{eq:ci-ic-interval} require differences in
verification success to shrink with drafting costs,
but allow differences in scope requirements to persist.
A stronger researcher can therefore continue to receive
more scope and less scrutiny than a weaker researcher,
even though both attain the same limiting verification
success. A vanishingly small difference in success can
sustain distinct assignments when the cost of additional
scope is sufficiently small.

Whether such differentiation survives depends on the
reviewer-time efficiency of the assignments. At a given
verification-success level, the editor can substitute
scope for scrutiny while preserving the success of
verification. As established in
Section~\ref{sec:scarcity}, the reviewer-efficient format
minimizes the reviewer time required to attain that
level. Because reviewer load depends on researcher type,
the most efficient combination of scope and scrutiny
may differ across types. Differentiated assignments
can therefore economize on reviewer time even when
verification success is equalized.

Formally, for any attainable verification-success level
$y$, let $x_i^E(y)$ denote the scope component of the
solution to
\[
\min_{x,r\geq 0}\ \ell(x, r, \theta_i)
\qquad
\text{subject to}\qquad y(x,r)=y.
\]
Thus $x_i^E(y)$ is the reviewer-efficient scope
requirement for type $i$ at success level $y$.
At the common limiting success level $y_T$, the editor
would like to assign each researcher this efficient
format. Incentive compatibility, however, requires
scope to be nondecreasing in researcher type.

With two types, suppose first that
$x_L^E(y_T)<x_H^E(y_T)$. The reviewer-efficient
formats respect the ordering required by incentive
compatibility. The editor can therefore retain
different combinations of scope and scrutiny for
the two types. The difference in verification success
needed to sustain these assignments vanishes with
drafting costs, while their difference in scope
converges to a positive value. Differentiation
survives because it saves reviewer time.

If instead $x_L^E(y_T)>x_H^E(y_T)$, the reviewer-efficient
formats conflict with incentive compatibility:
the weaker researcher would be assigned more scope
than the stronger researcher. The editor must sacrifice
some reviewer-time efficiency to respect the required
ordering. Under strict convexity of reviewer load
along the iso-success curve, the least costly ordered
allocation assigns the two types a common format.
Separating their scope requirements from this common
format entails a positive marginal reviewer-time cost,
while the incentive benefit of separation vanishes
as drafting costs fall. At sufficiently advanced
technology, pooling is therefore optimal.

The next theorem characterizes these two possibilities.
Equalization of verification success can coexist with
persistent differentiation when the reviewer-efficient
formats respect incentive-compatible sorting. Pooling
emerges when the two requirements conflict.

\begin{theorem}[Pooling and persistent differentiation]
\label{thm:ci-pooling}
Maintain the assumptions of Theorem~\ref{thm:ci-equalization}, and suppose there are two types. Suppose that, for each type, reviewer load at a fixed verification-success level is strictly convex in scope and attains an interior minimum. At fixed $T\in\mathcal T$:
\begin{enumerate}[(i)]
\item If $x_L^E(y_T)>x_H^E(y_T)$, the editor pools the two types at every sufficiently advanced production technology.
\item If $x_L^E(y_T)<x_H^E(y_T)$, their assignments remain distinct, with
\[
 x_H^*(T,\tau)-x_L^*(T,\tau)\longrightarrow x_H^E(y_T)-x_L^E(y_T)>0.
\]
\end{enumerate}
\end{theorem}

\hyperlink{proof_thm:ci-pooling}{The proof is in the appendix on page \pageref{proof_thm:ci-pooling}.}  The difference between the two cases is the reviewer-time value of differentiation. When the efficient formats conflict with sorting, creating a scope gap moves both researchers away from the formats that minimize reviewer time. At a finite technology, the gap can still sustain a difference in verification success, but~\eqref{eq:ci-ic-interval} shows that this incentive benefit shrinks with marginal drafting costs. Eventually the extra reviewer time costs more than the screening benefit, and the editor pools the types exactly. When the efficient formats respect sorting, the scope gap serves a productive purpose even as its signaling role disappears. The editor keeps differentiated scope and review assignments because they save reviewer time at the common success level.

\section{Reviewer Supply and the Verification Bottleneck}
\label{sec:bottleneck}

Section~\ref{sec:reorganization} held reviewer capacity fixed. Reviewers are volunteers. They decide how much time to supply in response to the expected scientific contribution of their assignments, and the editor allocates the resulting capacity by solving~\eqref{eq:editor-problem}. We first characterize equilibrium and establish existence, then show how the decline of the frontier translates into reviewer capacity and verified knowledge, when that decline is strict, and how its long-run course depends on verification technology.

\subsection{Reviewer participation and equilibrium}
\label{subsec:reviewer-equilibrium}

Reviewers enter before learning which submissions they will evaluate, so the return to supplying capacity is expected impact $I$, the verified value produced per unit of capacity. Taking $I$ as given, they solve
\begin{equation}\label{eq:reviewer-problem}
 \max_{T\geq0}\ \phi IT-C(T;\tau),
\end{equation}
whose solution satisfies $C_T(T)=\phi I$. In equilibrium the impact reviewers anticipate is the impact the editor delivers, $I=V^*(T,\tau)/T$, so positive equilibrium capacities solve
\begin{equation}\label{eq:equilibrium-condition}
 C_T(T)=\phi\frac{V^*(T,\tau)}{T}.
\end{equation}
Two features of this condition drive what follows. First, the opportunity cost of reviewing does not depend on production technology. AI makes writing a paper cheaper, but not the hours a senior researcher spends checking one, so any change in reviewer supply works through the scientific return to reviewing. Second, reviewers respond to the average contribution of an hour, while the editor values capacity at its marginal contribution $\gamma$. Reviewers supply more time when the editor uses the time she already has productively, and less when she does not.

Define the residual
\begin{equation}\label{eq:reviewer-residual}
 F(T;\tau)=C_T(T)-\phi\frac{V^*(T,\tau)}{T}.
\end{equation}
Reviewers wish to supply more capacity when $F<0$ and less when $F>0$. The editor's value function is nondecreasing and right-continuous in capacity. A positive pooled allocation produces verified value at least proportional to small capacities, so the residual is negative near zero; bounded verified value and Assumption~\ref{ass:C} make it positive at large capacities; and it has no upward jumps as capacity increases.

\begin{proposition}[Existence of peer-review equilibrium]\label{prop:existence}
Under Assumptions~\ref{ass:v}--\ref{ass:C}, a peer-review equilibrium exists for every production technology $\tau$, and the set of positive equilibrium reviewer capacities has a greatest element.
\end{proposition}

\hyperlink{proof_prop:existence}{We prove Proposition~\ref{prop:existence} in the appendix on page \pageref{proof_prop:existence}  using a compactification and crossing argument.}  Equilibrium can be multiple, because optimized impact can vary nonmonotonically with capacity. Write $\overline T(\tau)$ for the greatest zero of~\eqref{eq:reviewer-residual} and $\overline V^{PR}(\tau)=V^*(\overline T(\tau),\tau)$, and define $\overline T^{FI}(\tau)$ and $\overline V^{FI}(\tau)$ in the same way with $V^{FI}$ in place of $V^*$. At any positive equilibrium,
\begin{equation}\label{eq:equilibrium-value-identity}
 V=\frac{TC_T(T)}{\phi},
\end{equation}
which is strictly increasing in $T$, so the equilibrium with the most reviewer capacity also has the most verified knowledge.

\subsection{The Verification Bottleneck}
\label{subsec:equilibrium-bottleneck}

Theorem~\ref{thm:fixed-capacity} provides the first step: at sufficiently advanced technology, the editor obtains weakly less verified knowledge from any fixed, equilibrium-relevant reviewer capacity. The envelope formula~\eqref{eq:envelope} explains the sign. At a regular optimum in this region, participation and downward incentive constraints between distinct bundles are slack. The remaining binding cross-bundle constraints concern weak researchers imitating stronger researchers, so the envelope formula reduces to
\begin{equation}\label{eq:bottleneck-envelope}
 V^*_{\tau}(T,\tau)
 =\sum_{i<j}\lambda_{ij}
   \int_{x_i}^{x_j}c_{x\tau}(s,\theta_i;\tau)\,ds\leq0.
\end{equation}
Here $x_i\leq x_j$ by incentive-compatible sorting, and $c_{x\tau}<0$: cheaper scope erodes the cost difference deterring upward imitation. Each binding constraint thus weakly lowers the editor's optimum. Theorem~\ref{thm:fixed-capacity} establishes the same comparison across changes of optimal menus, including points where the envelope derivative is unavailable.

The second step is reviewer participation. The marginal opportunity cost $C_T(T)$ in~\eqref{eq:equilibrium-condition} is unchanged by production technology, while the expected contribution of an hour, $V^*(T,\tau)/T$, falls at each fixed capacity. Equivalently, differentiating~\eqref{eq:reviewer-residual} where the frontier is smooth gives
\[
 F_{\tau}(T;\tau)=-\frac{\phi}{T}V^*_{\tau}(T,\tau)\geq0.
\]
On a locally stable equilibrium branch, where $F_T>0$, implicit differentiation yields $dT/d\tau=-F_{\tau}/F_T\leq0$. This withdrawal of reviewer time compounds the decline in the editor's productivity. The global argument uses the greatest equilibrium: since $F$ shifts upward at every relevant capacity, no root can appear above the previous greatest root. Finally, verified knowledge falls with capacity along the equilibrium identity~\eqref{eq:equilibrium-value-identity}. Under full information the editor's frontier rises instead, shifting reviewer incentives in the opposite direction.

\begin{theorem}[Verification bottleneck]\label{thm:bottleneck}
Under Assumptions~\ref{ass:v}--\ref{ass:C}, there exists a finite $\bar\tau$ such that, for every $\tau_2>\tau_1>\bar\tau$,
\[
 \overline T(\tau_2)\leq\overline T(\tau_1),
 \qquad
 \overline V^{PR}(\tau_2)\leq\overline V^{PR}(\tau_1).
\]
Under full information, equilibrium reviewer capacity and verified knowledge are weakly increasing in production technology.
\end{theorem}

\hyperlink{proof_thm:bottleneck}{The proof is in the appendix on page \pageref{proof_thm:bottleneck}.} The theorem combines a decline in the value produced by a fixed stock of reviewer time with a reduction in the stock itself. As cheaper drafting forces the editor to compress differences between assignments or to devote resources to deterring imitation, reviewers anticipate less scientific impact from an hour's scrutiny and supply fewer hours. Equilibrium verified knowledge declines through both channels. The comparison extends to multiple equilibria because it tracks their greatest capacity rather than assuming a smooth equilibrium path.

Along a differentiable, locally stable branch of equilibria, differentiating~\eqref{eq:equilibrium-condition} gives
\begin{equation}\label{eq:amplification}
 \frac{dV}{d\tau}=V_\tau^*\,\frac{1+\varepsilon_C}{1+\varepsilon_C-\varepsilon_V},
 \qquad
 \varepsilon_V=\frac{TV_T^*}{V^*},
 \qquad
 \varepsilon_C=\frac{TC_{TT}}{C_T},
\end{equation}
where stability means $1+\varepsilon_C>\varepsilon_V$. Reviewer withdrawal multiplies the fixed-capacity loss by a factor of at least one, larger when verified value responds more to capacity and when marginal reviewing cost is flatter; with quadratic reviewing costs the factor is $2/(2-\varepsilon_V)$.

Under full information, cheaper drafting relaxes participation without creating an imitation problem. The full-information equilibrium therefore moves in the opposite direction, eventually becoming constant once participation constraints cease to bind (Lemma~\ref{lem:fi-frontier}); its reviewer capacity and verified knowledge are weakly greater than under private information at every technology. The contrast isolates the loss of information conveyed by scope requirements, rather than a change in the physical productivity or opportunity cost of reviewer time.

The editor's optimized frontier gives the highest verified value attainable through feasible allocations of scrutiny. Editorial practices such as desk rejection, scope limits, and referee triage determine how closely an institution approaches that frontier. Theorem~\ref{thm:bottleneck} establishes that the frontier itself weakly declines at sufficiently advanced production technology. This mechanism provides a distinct channel of peer-review strain alongside the effects of growing submission volumes studied by \citet{BergstromGross2026,Atasu2026}. Even with a fixed population of researchers, cheaper production can reduce the expected scientific contribution of reviewing, leading to lower voluntary reviewer capacity and less verified knowledge.

\subsection{Strict bottleneck under separability}
\label{subsec:strict}

The weak inequalities in Theorem~\ref{thm:bottleneck} permit the decline to stop. One possibility is that all types eventually pool, so further reductions in drafting costs leave their assignments unchanged. Theorem~\ref{thm:ci-pooling} shows why this can happen. The alternative arises when researcher quality gives types different comparative advantages in using scope rather than scrutiny. Then differentiated verification formats can remain productive even as differences in verification success vanish.

Under separable verified value~\eqref{eq:separable}, moving along a curve of constant verification success requires $dr/dx=-y_x/y_r$. For type $i$, the resulting marginal change in reviewer load is
\begin{equation}\label{eq:format-review-cost}
 D_i(x,r):=\ell_x(x, r, \theta_i)
 -\ell_r(x, r, \theta_i)\frac{y_x(x,r)}{y_r(x,r)}.
\end{equation}
A lower $D_i$ means a greater reviewer-time advantage from making the verification format more scope-intensive. The following condition makes that advantage strictly greater for higher-quality researchers and gives a strict bottleneck for any finite number of types.

\begin{theorem}[Strict verification bottleneck]\label{thm:strict-M}
Suppose Assumptions~\ref{ass:v}--\ref{ass:C} hold, verified value is separable as in~\eqref{eq:separable}, and $M\geq2$. Suppose, at all positive feasible bundles and qualities,
\[
 \ell_{r\theta}(x, r, \theta)<0,
 \qquad
 \frac{\ell_{x\theta}(x, r, \theta)}{\ell_{r\theta}(x, r, \theta)}
 >\frac{y_x(x,r)}{y_r(x,r)}.
\]
Then there exists a finite $\tau_s$ such that, for every $\tau_2>\tau_1>\tau_s$,
\[
 \overline T(\tau_2)<\overline T(\tau_1),
 \qquad
 \overline V^{PR}(\tau_2)<\overline V^{PR}(\tau_1).
\]
Moreover, $V^*(T,\tau_2)<V^*(T,\tau_1)$ at every equilibrium-relevant capacity $T\in[T_-,T_+]$.
\end{theorem}

\hyperlink{proof_thm:strict-M}{The proof is in the appendix on page \pageref{proof_thm:strict-M}.}  Direct differentiation shows that the exponential-bandit technology of Section~\ref{subsec:experimentation} satisfies the additional conditions of the theorem for $\alpha\geq2$ under the assumptions of Proposition~\ref{prop:microfoundation}.

To understand the result, begin with a pooled menu. At the reviewer-efficient pooled bundle, the population-weighted average of the format costs $D_i$ is zero. The strict ordering~\eqref{eq:strict-format-advantage} then makes $D_M<0$. Shifting the strongest researchers toward higher scope and less scrutiny saves reviewer time at unchanged verification success. To make that departure incentive compatible, the editor must give the stronger group a small advantage in verification success. The necessary advantage shrinks with marginal drafting costs, while the reviewer-time saving remains. Full pooling is therefore suboptimal once production is sufficiently cheap.

The remaining question is whether the editor can differentiate researchers without paying an incentive cost. At high technology, the comparative-verification-advantage assumption orders the marginal product of reviewer time strictly across distinct groups in an optimal menu. If every incentive constraint between those groups had zero shadow value, the editor's first-order conditions would instead equalize their marginal products. This contradiction establishes a positive shadow value on at least one upward imitation constraint, which makes the fixed-capacity loss strict. Reviewer withdrawal then makes both equilibrium inequalities strict. The argument permits some adjacent types to pool; it requires only that the editor does not pool everyone.

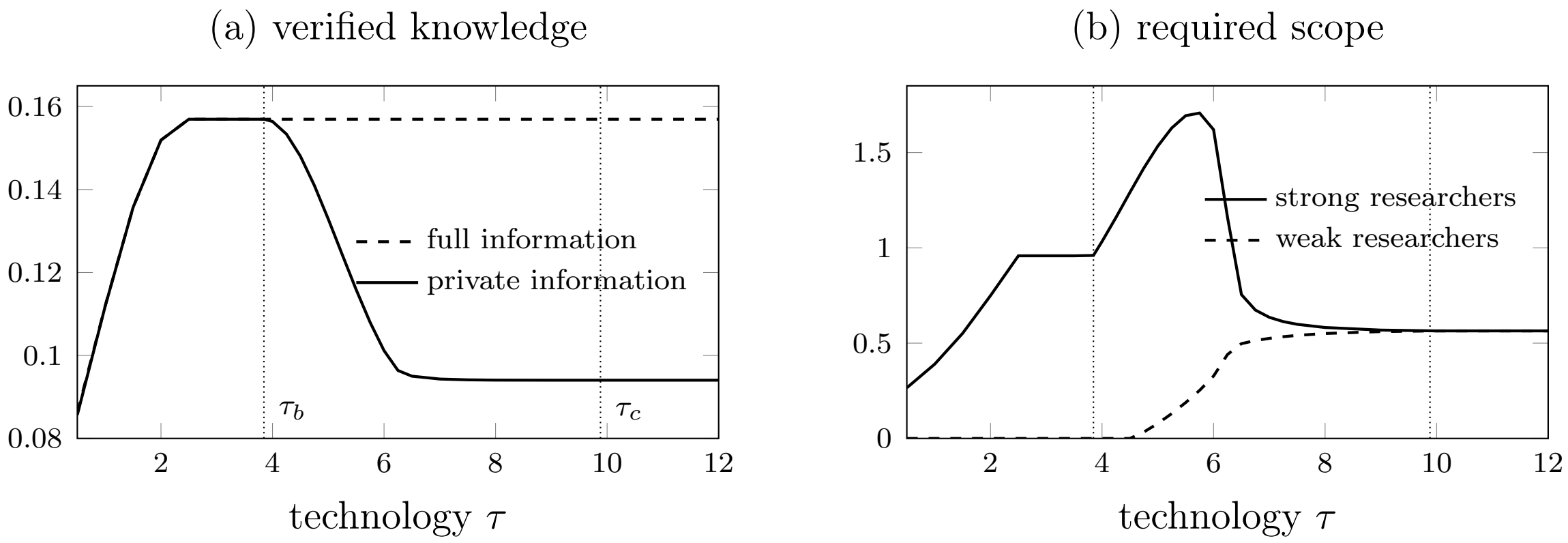
\begin{figure}[t]
\centering
\begin{minipage}{0.49\textwidth}
\centering
\begin{tikzpicture}
\begin{axis}[
 width=\textwidth, height=5.0cm,
 xmin=0.5, xmax=12, ymin=0.08, ymax=0.165,
 xlabel={technology $\tau$}, title={(a) verified knowledge},
 title style={font=\small}, tick label style={font=\scriptsize}, label style={font=\small},
 yticklabel style={/pgf/number format/fixed,/pgf/number format/precision=2},
 legend style={font=\scriptsize,draw=none,fill=none,at={(0.98,0.5)},anchor=east},
 legend cell align=left]
\addplot[thick,dashed] coordinates {(0.5,0.08623)(1,0.11197)(1.5,0.13569)(2,0.15191)(2.5,0.15695)(12,0.15695)};
\addlegendentry{full information}
\addplot[thick] coordinates {(0.5,0.08567)(1,0.11188)(1.5,0.13569)(2,0.15191)(2.5,0.15695)(3,0.15695)(3.5,0.15695)(3.846,0.15695)(4,0.15641)(4.25,0.15337)(4.5,0.14804)(4.75,0.14100)(5,0.13288)(5.25,0.12433)(5.5,0.11587)(5.75,0.10799)(6,0.10117)(6.25,0.09638)(6.5,0.09501)(7,0.09432)(7.5,0.09413)(8,0.09406)(9,0.09404)(10,0.09404)(12,0.09404)};
\addlegendentry{private information}
\draw[densely dotted] (axis cs:3.846,0.08) -- (axis cs:3.846,0.165);
\draw[densely dotted] (axis cs:9.881,0.08) -- (axis cs:9.881,0.165);
\node[font=\scriptsize,anchor=south west] at (axis cs:3.846,0.081) {$\tau_b$};
\node[font=\scriptsize,anchor=south west] at (axis cs:9.881,0.081) {$\tau_c$};
\end{axis}
\end{tikzpicture}
\end{minipage}\hfill
\begin{minipage}{0.49\textwidth}
\centering
\begin{tikzpicture}
\begin{axis}[
 width=\textwidth, height=5.0cm,
 xmin=0.5, xmax=12, ymin=0, ymax=1.85,
 xlabel={technology $\tau$}, title={(b) required scope},
 title style={font=\small}, tick label style={font=\scriptsize}, label style={font=\small},
 legend style={font=\scriptsize,draw=none,fill=none,at={(0.98,0.62)},anchor=east},
 legend cell align=left]
\addplot[thick] coordinates {(0.5,0.2657)(1,0.3900)(1.5,0.5514)(2,0.7492)(2.5,0.9582)(3,0.9582)(3.5,0.9582)(3.846,0.9602)(4,1.0332)(4.25,1.1593)(4.5,1.2907)(4.75,1.4186)(5,1.5336)(5.25,1.6288)(5.5,1.6934)(5.75,1.7072)(6,1.6202)(6.25,1.1602)(6.5,0.7547)(6.75,0.6736)(7,0.6354)(7.25,0.6131)(7.5,0.5988)(8,0.5821)(9,0.5685)(10,0.5643)(12,0.5643)};
\addlegendentry{strong researchers}
\addplot[thick,dashed] coordinates {(0.5,0)(4.5,0)(4.75,0.0337)(5,0.0784)(5.25,0.1305)(5.5,0.1883)(5.75,0.2524)(6,0.3273)(6.25,0.4405)(6.5,0.4973)(6.75,0.5138)(7,0.5251)(7.25,0.5338)(7.5,0.5407)(8,0.5506)(9,0.5607)(10,0.5643)(12,0.5643)};
\addlegendentry{weak researchers}
\draw[densely dotted] (axis cs:3.846,0) -- (axis cs:3.846,1.85);
\draw[densely dotted] (axis cs:9.881,0) -- (axis cs:9.881,1.85);
\end{axis}
\end{tikzpicture}
\end{minipage}
\caption{The equilibrium path in the exponential bandit technology for $\alpha=0$. Panel (a): verified knowledge at the maximal equilibrium. Panel (b): scope assigned to each type under private information. Along this computed path the editor first escalates the strong bundle's scope requirement and then compresses the two bundles until she pools them.}
\label{fig:path}
\end{figure}

The additional comparative-advantage restriction identifies a regime in which research quality affects the relative reviewer-time efficiency of scope and scrutiny. Higher-quality researchers have a comparative advantage in scope-intensive verification formats, so differentiated assignments retain productive value as drafting costs vanish. When the reviewer-efficient formats instead conflict with incentive-compatible sorting, the editor may eventually pool researchers, allowing the decline to reach a plateau. This occurs in the exponential-bandit specification with $\alpha=0$, whose equilibrium path is illustrated in Figure~\ref{fig:path}. The long-run consequences of cheaper research production therefore depend on how researcher quality affects the relative reviewer-time efficiency of scope and scrutiny.

\section{Welfare and Institutional Responses}
\label{sec:welfare}

The bottleneck concerns verified knowledge, the institutional output of peer review. Its welfare consequences also depend on the resources used to produce research and on the opportunity cost of reviewing, and institutional remedies act on different parts of the mechanism. We first separate these margins, then examine changes to the information, verification technology, and instruments available for allocating scrutiny. The comparisons separate the restrictions intrinsic to scarce verification from those that arise from the way peer review organizes it.

\subsection{Verified knowledge and resource-adjusted surplus}
\label{subsec:welfare-accounting}

The editor maximizes verified knowledge. A welfare comparison would also count the real costs of producing research and of reviewers' time. Normalize reviewer cost by $\widetilde C(T)=C(T;\tau)-C(0;\tau)$ and define resource-adjusted surplus
\begin{equation}\label{eq:resource-surplus}
 S(\varrho,T;\tau)
 :=\sum_{i=1}^M f_i\bigl[v(x_i, r_i, \theta_i)-c(x_i,\theta_i;\tau)\bigr]
 -\widetilde C(T),
 \qquad L(\varrho)\leq T,
\end{equation}
the scientific contribution established through review net of the real costs of production and reviewer time. Reviewers' concern for the impact of their assignments determines how much time they supply, and verified knowledge enters surplus once. Drafting costs, by contrast, enter the editor's problem only through incentive and participation constraints, so peer review solves a different problem from maximizing surplus. To compare the two on the margin that drives the bottleneck, reviewer supply, let $B^*(T,\tau)$ be the greatest value of $\sum_i f_i[v(x_i, r_i, \theta_i)-c(x_i,\theta_i;\tau)]$ over incentive-compatible, individually rational menus with load at most $T$. A planner able to mobilize capacity, subject to the same information and instruments, would set $B_T^*(T,\tau)=C_T(T)$, whereas reviewers set $C_T(T)=\phi V^*(T,\tau)/T$. The wedge between the two margins,
\begin{equation}\label{eq:reviewer-supply-wedge}
 B_T^*(T,\tau)-\phi\frac{V^*(T,\tau)}{T},
\end{equation}
compares the marginal contribution of capacity to net research benefits with the average contribution a reviewer values.

The wedge can take either sign. Its sign depends on the productivity of additional reviewer time, the drafting costs of the allocations it supports, and the weight reviewers place on scientific impact. Voluntary supply can therefore fall short of the surplus benchmark or exceed it, and Theorem~\ref{thm:bottleneck} concerns the direction in which reviewer capacity and verified knowledge move, not whether their levels are efficient. Cheaper production also releases resources, so a decline in verified knowledge can coexist with savings in production costs, and total surplus can move either way. With observable quality, by contrast, maximized surplus weakly rises with technology, because the feasible set expands and every fixed allocation becomes cheaper. The difference is the welfare counterpart of the screening mechanism: under private information, cheaper drafting also makes high-scope assignments more attractive to weaker researchers, and implementing them becomes more expensive.

\subsection{Information and the productivity of verification}
\label{subsec:institutions-information}

Two institutional changes expand what reviewer time can achieve. First, a better verification technology reduces the reviewer time needed for a given depth of scrutiny. Let $\eta$ index verification technology, with load $\ell(x, r, \theta;\eta)$ satisfying $\ell(x, 0, \theta;\eta)=0$ and
\begin{equation}\label{eq:verification-improvement}
 \ell_{r\eta}(x, r, \theta;\eta)<0
 \qquad (x,r>0),
\end{equation}
so that any assignment uses less reviewer time while researchers' preferences are unchanged. Second, better preliminary information lets the editor distinguish researchers before assigning scrutiny. Suppose an exogenous, nonmanipulable signal reveals a cell of a partition $\mathcal P$ of the types before scope is chosen, so that incentive compatibility applies only within cells. The first change acts on the capacity constraint and the second on the incentive constraints.

\begin{proposition}[Information and verification technology]
\label{prop:institutional-improvements}
Fix production technology $\tau$, and suppose each environment compared satisfies Assumptions~\ref{ass:v}--\ref{ass:C}. A better verification technology, a finer partition, or both weakly raise the verified value attainable from every reviewer capacity, and they weakly raise the greatest equilibrium reviewer capacity and verified knowledge.
\end{proposition}

The proof is in the appendix on page \pageref{proof_prop:institutional-improvements}. Each change leaves every previously feasible menu feasible, so the editor can do at least as well with given capacity. Reviewers anticipate the higher return and supply at least as much time, so the gain carries over to equilibrium, through the same participation response that amplifies the bottleneck. The two changes act on different parts of the mechanism. Verification technology economizes on the scarce input. Automated checks of calculations, code, replication, or formal proofs act on this margin and free expert attention for judgments about novelty and significance. Preliminary information changes which allocations are incentive compatible, so the editor can direct scrutiny by quality without costly scope distinctions. Desk screening that is informative about quality beyond what scope reveals, evidence about research methods, or reliable pre-review signals act on this margin. By Theorem~\ref{thm:fixed-capacity}, the value of such information at given capacity grows with production technology beyond $\bar\tau$, so it becomes more valuable precisely as cheaper production erodes screening through scope.

\subsection{Submission fees}
\label{subsec:submissionfees}

A fee conditioned on scope is a deterrent whose value does not depend on drafting costs, so it can replace the drafting-cost differences whose erosion drives the reviewer cost of separation in Theorem~\ref{thm:cost-dependent-separation}. Suppose the editor can set a nonnegative schedule $s(x)$ with $s(0)=0$, so that a researcher choosing $(x,r)$ obtains $v(x, r, \theta)-c(x,\theta;\tau)-s(x)$, with fee revenue excluded from the objective. For two types and a proposed menu with $x_L<x_H$, let $U_i^j=v(x_j, r_j, \theta_i)-c(x_j,\theta_i;\tau)$ and $\Delta_i=U_i^H-U_i^L$. If the bundles are individually rational under full information, incentive compatibility requires
\begin{equation}\label{eq:fee-screening-interval}
 \Delta_L\leq s(x_H)-s(x_L)\leq\Delta_H,
\end{equation}
while participation bounds both fees.

\begin{proposition}[Implementation with scope-contingent fees]
\label{prop:submissionfees}
Suppose $0<x_L<x_H$ and the proposed allocation satisfies participation under full information. It can be implemented by nonnegative fees on the two assigned scope if and only if
\begin{equation}\label{eq:fee-implementability}
 \max\{\Delta_L,-U_L^L\}
 \leq\min\{\Delta_H,U_H^H\}.
\end{equation}
If instead $z_L=(0,0)$ and $x_H>0$, implementation is possible if and only if
\begin{equation}\label{eq:fee-null-implementability}
 \max\{0,U_L^H\}\leq U_H^H.
\end{equation}
Under Assumption~\ref{ass:u}, the latter condition holds whenever the high-type bundle is individually rational.
\end{proposition}

The proof is in the appendix on page \pageref{prop:submissionfees}. The conditions ask only that the fee difference needed to keep the weak type out be compatible with both types' participation. A common flat fee on two positive-scope bundles cancels from~\eqref{eq:fee-screening-interval}: it affects entry, not screening. In the running example, the full-information allocation, which excludes the weak type and assigns the strong type $z_H$ with success index $y_H$, is implemented by any fee with $\max\{0,a_Ly_H-4e^{-\tau}x_H^2\}\leq s(x_H)\leq a_Hy_H-4e^{-\tau}x_H^2$, a nonempty interval whenever the strong type participates. Since the fee-augmented frontier is bounded by the full-information frontier and coincides with it at an implemented optimum, scope-contingent fees restore the full-information equilibrium at advanced technologies in the example. Journal fees, by contrast, are typically flat. In a census of 500 economics journals, 22.6 percent charge a submission fee, with a median of \$143 among those that do \citep{Fourie2026}, and their burden varies with researchers' resources \citep{OldfordFisetArmenakyan2023}.

\subsection{Pricing reviewer time}
\label{subsec:pricing}

Fees act on the incentive problem. A broader change is to act on the scarcity of reviewer time itself, by letting researchers buy verification. Suppose the reviewer time used on a submission is contractible and priced at $p\geq0$. Researchers choose their own bundles to maximize $v(x, r, \theta_i)-c(x,\theta_i;\tau)-p\ell(x,\theta_i,r)$, reviewers choose $T$ to maximize $pT-\widetilde C(T)$, and the market clears when $L(\varrho)=T$. Consider the surplus benchmark with unconstrained researcher choices,
\begin{equation}\label{eq:market-surplus}
 S^M(\tau)
 :=\max_{\varrho,T\geq0}
 \left\{\sum_i f_i\bigl[v(x_i, r_i, \theta_i)-c(x_i,\theta_i;\tau)\bigr]
 -\widetilde C(T)
 : L(\varrho)\leq T\right\}.
\end{equation}

\begin{proposition}[Competitive pricing of verification]
\label{prop:pricing}
Every competitive verification equilibrium, if one exists, attains $S^M(\tau)$. Conversely, a surplus-maximizing allocation can be decentralized at a nonnegative price $p$ whenever it globally maximizes the price-adjusted objective
\begin{equation}\label{eq:pricing-lagrangian}
 \mathcal J_p(\varrho,T;\tau)
 :=\sum_i f_i\bigl[v(x_i, r_i, \theta_i)-c(x_i,\theta_i;\tau)
 -p\ell(x_i,\theta_i,r_i)\bigr]
 +pT-\widetilde C(T).
\end{equation}
For every $\tau_2>\tau_1$, $S^M(\tau_2)\geq S^M(\tau_1)$, with strict inequality whenever a surplus-maximizing allocation at $\tau_1$ assigns positive scope to at least one type.
\end{proposition}

The proof is in the appendix on page \pageref{proof_prop:pricing}. When researchers pay for the reviewer time they use, private information no longer matters for the allocation, because each researcher chooses her own bundle and bears its full cost in reviewer time. Cheaper production then weakly raises maximized surplus, even though verified knowledge under peer review can fall. The benchmark shows that the bottleneck stems from allocating scarce verification without prices, through screening on scope. Contractibility is demanding, however: payment by the hour or per report rewards neither care nor thoroughness, so contracting on the quality of review adds an incentive problem of its own.

\section{Conclusion}

Advances in research production technology change how scarce scientific attention is allocated. When researcher quality is privately known, research scope serves as a signal of quality as well as a source of potential scientific value. We show that screening through scope distorts both research decisions and the allocation of scrutiny, directing reviewer time away from its most productive uses. As production becomes cheaper, maintaining distinctions between researchers requires increasingly costly scope demands, because weaker researchers can more readily imitate stronger ones. The editor must consequently adjust how research is screened and verified, even when editorial decisions are optimal.

These forces can create a verification bottleneck in which both verified knowledge and the voluntary supply of reviewer time decline as research production becomes cheaper. Beyond a technological threshold, screening constraints reduce the knowledge attainable from a given reviewer capacity. The resulting decline in the expected scientific contribution of reviewing induces reviewers to withdraw time, further reducing equilibrium verified knowledge. The deterioration can persist when different researcher types retain a comparative advantage in different verification formats, or reach a plateau when pooling becomes optimal. When researcher quality is observable, cheaper production instead weakly increases both verified knowledge and reviewer capacity. The bottleneck therefore arises from asymmetric information, even with unchanged verification technology and reviewer opportunity costs.

The findings have implications for how scientific evaluation is organized. More informative preliminary screening can reduce reliance on costly scope requirements, while improvements in verification technology allow scarce reviewer time to establish more knowledge. Scope-contingent fees and markets for reviewer time offer alternative ways of aligning research and verification incentives. The distinction between verified knowledge and economic surplus also matters: cheaper research can reduce resource costs even when the knowledge established through peer review declines. More broadly, the same problem can arise wherever institutions must evaluate increasingly inexpensive submissions from privately informed producers. As research becomes easier to produce, the ability to distinguish valuable contributions and establish their merits becomes increasingly important to the productivity of scientific institutions.

\newpage
\begin{appendix}
\section{Proofs}\label{sec:proofs}

Three preliminary lemmas supply the regularity and incentive properties used in several results. The proofs proceed from pairwise incentive constraints. Where types share an assignment, summing their first-order conditions cancels the internal IC multipliers. We explain each construction as it arises and give the equations needed for the comparison.

\subsection{Three supporting lemmas}

The screening, fixed-capacity, equalization, and bottleneck proofs pass to the limit as drafting costs vanish. For those limits and their first-order conditions, we need optimal bundles to stay in a common compact interior set, with slack participation and $u_x>0$. We also need one capacity interval containing advanced equilibria. The next lemma supplies both facts.

\begin{lemma}[Uniformly bounded optimal menus]\label{lem:uniform-optima}
There are $0<T_-<T_+<\infty$, $\tau_0<\infty$, and a compact set
$K\subset(0,\infty)^2$ such that all equilibrium reviewer capacities at
$\tau\geq\tau_0$ lie in $\mathcal T:=[T_-,T_+]$, and every optimal menu for
$T\in\mathcal T$, $\tau\geq\tau_0$ assigns each type a bundle in $K$.
On these menus, participation is slack and $u_x>0$. The same menu conclusions
hold at any fixed $T>0$, with constants depending on $T$.
\end{lemma}

\begin{proof}
Verification-load coercivity bounds a bundle's inputs when its verified value is bounded away from zero. The issue is to establish that positive-value bound for \emph{every} optimal assignment. We first bound equilibrium capacity, then find one valuable bundle, and use pairwise IC to transfer its positive payoff to all types. Coercivity finishes the compactness argument.

\emph{Step 1: Locate equilibrium capacity in a common interval.}
Fix $r_0>0$ and pool all types at $(x,r_0)$ for small $x>0$. Assumption~\ref{ass:u} gives positive marginal private utility near zero, $u_x>0$ as $x\downarrow0$. Since $v=u+c$ and $c\geq0$, verified value satisfies $v(x, r_0, \theta_i)\geq a_i x$ for some $a_i>0$. Assumption~\ref{ass:ell} gives $\ell(0, r_0, \theta_i)=0$ and finite $\ell_x$ there, hence $\ell(x, r_0, \theta_i)\leq b_i x$ for some finite $b_i$. Pooling ensures IC, positive utility ensures participation, and cheaper drafting preserves participation at higher $\tau$.

With finitely many types, choose common $a,b>0$: pooled verified value is at least $ax$, and load at most $bx$. Setting $x=T/b$ for small capacity $T$ gives
\[
\frac{V^*(T,\tau)}{T}\geq\frac ab>0.
\]
Reviewers' marginal cost satisfies $C_T(T)\to0$ as $T\downarrow0$, while impact $V^*/T\geq a/b$. Thus the supply residual $F=C_T-\phi V^*/T$ is negative near zero. For large $T$, bounded verified value $V^*\leq\bar V$ and $C_T(T)\to\infty$ make $F$ positive. These uniform sign bounds imply that every positive equilibrium capacity, whenever it exists, lies in a common interval $\mathcal T=[T_-,T_+]\subset(0,\infty)$. Proposition~\ref{prop:existence} establishes existence.

\emph{Step 2: Obtain one valuable, bounded bundle.}
Choose a positive pooled menu with load below $T_-$. It remains feasible as $\tau$ rises and delivers some $\widehat V>0$. Every optimum on $\mathcal T$ therefore has $V^*\geq\widehat V$. Since $V^*=\sum_i f_i v(z_i,\theta_i)$, some type $j$ satisfies $v(z_j,\theta_j)\geq\widehat V$; its individual load is at most $T_+/f_j$:
\[
v(z_j,\theta_j)\geq\widehat V,\qquad
\ell(z_j,\theta_j)\leq T_+/f_j.
\]
Assumption~\ref{ass:ell} says bounded load and positive verified value bound both inputs. Also $v=0$ on either axis, so $z_j$ stays in a compact subset of the positive quadrant. At the same bundle every type has $v(z_j,\theta_i)>0$; finiteness of the type set and continuity give a common bound $v(z_j,\theta_i)\geq\alpha>0$. Every type can obtain this value by imitating $j$, up to drafting costs.

\emph{Step 3: Use pairwise imitation to bound all bundles.}
The constraint against type $i$ choosing type $j$'s assignment gives
\[
 u(z_i,\theta_i;\tau)\geq u(z_j,\theta_i;\tau)
 =v(z_j,\theta_i)-c(x_j,\theta_i;\tau).
\]
Drafting costs vanish uniformly on the bounded scope choices of Step 2, so for large $\tau$,
\[
u(z_i,\theta_i;\tau)\geq\alpha/2,
\qquad v(z_i,\theta_i)\geq\alpha/2\quad\text{for every }i.
\]
Now each type has $v_i\geq\alpha/2$ and load $\ell_i\leq T_+/f_i$. Load coercivity again bounds $x_i,r_i$, and $v=0$ on the axes keeps both positive. Thus all optimal bundles lie in one compact interior set $K$, with slack participation. On $K$, $v_x>0$ is bounded away from zero; Assumption~\ref{ass:c} gives $c_x\to0$ uniformly, hence $u_x=v_x-c_x>0$ for large $\tau$.

For any fixed $T>0$, begin Step 2 with a positive pooled menu using less than $T$. Steps 2--3 then establish the fixed-capacity version directly.
\end{proof}

The screening proposition and the fixed-capacity and strict-bottleneck theorems use the fact that, at advanced technology, the binding constraints between distinct assignments are upward IC constraints. The next lemma supplies this orientation by proving downward IC strictly slack.

\begin{lemma}[Slack downward incentive constraints]\label{lem:downward}
Under Assumptions~\ref{ass:v}--\ref{ass:q}, for every fixed
$T>0$ there is a finite threshold beyond which every downward incentive
constraint between distinct optimal bundles is strictly slack.
If Assumption~\ref{ass:C} also holds, a common threshold
$\tau_{\mathrm{IC}}$ works uniformly for all
$T\in\mathcal T$ from Lemma~\ref{lem:uniform-optima}.
\end{lemma}

\begin{proof}
Suppose a higher type is indifferent between her assigned bundle and a lower-scope bundle. As $c\to0$, the pair approaches the same iso-value curve for that type. Comparative verification advantage, Assumption~\ref{ass:q}, then ranks marginal reviewer productivity upward across the pair; the editor's first-order conditions under binding downward IC rank it downward. The proof makes this contradiction precise, using block sums only where types share a bundle.

\emph{Step 1: Reduce the incentive system to adjacent comparisons.}
By Lemma~\ref{lem:uniform-optima}, assigned bundles are interior and $u_x>0$; Assumption~\ref{ass:v} gives $u_r=v_r>0$. Strict single crossing, $\partial_\theta(u_x/u_r)>0$, and pairwise IC imply $x_1\leq\cdots\leq x_M$. Equal scope imply equal depths because $u_r>0$. Types receiving the same bundle therefore form consecutive pooling groups.

For three distinct ordered bundles $z^a,z^b,z^c$, a type in the middle group prefers $z^b$ to $z^c$. Single crossing makes every lower type strictly prefer $z^b$ to $z^c$; since those types also prefer $z^a$ to $z^b$, their IC against $z^c$ is slack. The downward comparison is symmetric. Thus binding cross-group IC links adjacent groups only. At each boundary the potential upward imitator is the highest type below it, and the potential downward imitator the lowest type above it; only one direction can bind.

\emph{Step 2: Justify the use of first-order multipliers.}
We use KKT multipliers below, so we verify MFCQ: one perturbation must strictly relax all active constraints. Let $B_1,\ldots,B_J$ be the consecutive pooling groups. Reduce the common review depth in each group by $\rho_k<0$. At each boundary, choose the relative reductions so the binding imitator strictly prefers her assigned group. For upward IC the condition is
\[
u_r(z^k,\theta_{\max B_k})\rho_k
-u_r(z^{k+1},\theta_{\max B_k})\rho_{k+1}>0;
\]
for a binding downward constraint it is
\[
u_r(z^{k+1},\theta_{\min B_{k+1}})\rho_{k+1}
-u_r(z^k,\theta_{\min B_{k+1}})\rho_k>0.
\]
Since $u_r>0$, the next $\rho_{k+1}<0$ can satisfy whichever inequality binds. Lower review depth also relaxes the capacity constraint, as $\ell_r>0$.

Within each pooled group, the IC constraints are equalities. Single crossing gives distinct slopes $m_i=u_x(z^k,\theta_i;\tau)/u_r(z^k,\theta_i;\tau)$. Add a small type-specific perturbation proportional to $e_i=(m_i,-m_i^2/2)$. Type $i$ strictly prefers its own perturbation to type $j$'s because
\[
\nabla u(z^k,\theta_i;\tau)\cdot(e_i-e_j)
=\tfrac12u_r(z^k,\theta_i;\tau)(m_i-m_j)^2>0.
\]
The square is positive for $i\ne j$. Scale these perturbations to preserve the cross-group and capacity slack already created. All active constraints improve, so MFCQ holds and KKT multipliers exist.

\emph{Step 3: Assume a downward pair remains binding.}
Suppose downward IC binds along $\tau_n\to\infty$ with $T_n\in\mathcal T$. Compactness gives convergent bundles; finiteness of types lets us fix the pooling partition and binding pattern along a subsequence. Select a maximal chain $B_a,\ldots,B_b$ connected by downward IC. At interior groups, multiplier terms enter from both sides; at the endpoints they have one sign. Let $\gamma_n$ be the capacity multiplier.

For any shared bundle $z^B$, define its aggregate marginal verified value and reviewer load in review depth by
\[
W_{B,r}=\sum_{i\in B}f_i v_r(z^B,\theta_i),\qquad
L_{B,r}=\sum_{i\in B}f_i\ell_r(z^B,\theta_i),\qquad
Q_B=\frac{W_{B,r}}{L_{B,r}}.
\]
Sum the review-depth FOCs for all types in $B$. Each within-group imitation multiplier enters twice with opposite signs and cancels, leaving
\begin{equation}\label{eq:downward-block-review}
W_{B,r}-\gamma_n L_{B,r}
=-\sum_{\substack{i\in B\\h\notin B}}\lambda_{ih}u_r(z^B,\theta_i)
+\sum_{\substack{i\notin B\\h\in B}}\lambda_{ih}u_r(z^B,\theta_i).
\end{equation}
At the lower endpoint of a downward chain, surviving IC terms give $W_{B_a,r}-\gamma_nL_{B_a,r}\geq0$; at the upper endpoint they give $W_{B_b,r}-\gamma_nL_{B_b,r}\leq0$. Dividing by $L_{B,r}>0$ yields
\begin{equation}\label{eq:downward-wrong-order}
Q_{B_a}\geq\gamma_n\geq Q_{B_b}>0.
\end{equation}
Each $Q_B$ is a positively weighted average of $q=v_r/\ell_r$. On the compact interior set, $0<\underline q\leq q\leq\overline q<\infty$, hence $0<\underline q\leq\gamma_n\leq\overline q$. These bounds allow division by $\gamma_n$ in Step 4.

\emph{Step 4: Analyze one binding downward pair.}
Take one binding downward pair across adjacent groups $B_k<B_{k+1}$, with $j=\min B_{k+1}$. Its IC condition is
\[
u(z^{k+1},\theta_j;\tau_n)=u(z^k,\theta_j;\tau_n)
\]
At the higher bundle, consider $dx_j=-\varepsilon$, $dr_j=t\varepsilon$. We choose $t$ so type $j$ likes the assignment with less scope that is more deeply reviewed, while higher types find it less attractive. Single crossing supplies the interval
\[
\frac{u_x(z^{k+1},\theta_j;\tau_n)}{u_r(z^{k+1},\theta_j;\tau_n)}
<t<
\frac{u_x(z^{k+1},\theta_{j+1};\tau_n)}{u_r(z^{k+1},\theta_{j+1};\tau_n)}
\]
when $j<M$, or simply $t$ above the first ratio when $j=M$. The lower bound on $t$ makes $du_j>0$; the upper bound gives $du_h<0$ for any higher type $h$, so those types are less inclined to imitate the modified entry. Lower types' upward IC is already slack (Step 1). Thus this is a feasible one-sided perturbation, including within a pooled upper group.

Optimality requires the first-order value of this feasible direction, net of reviewer time priced at $\gamma_n$, to be nonpositive. At type $j$'s upper bundle, write $m_j=v_x/v_r$ and $k_j=\ell_x/\ell_r$; then
\[
 v_r(z^{k+1},\theta_j)(t-m_j)
-\gamma_n\ell_r(z^{k+1},\theta_j)(t-k_j)\leq0,
\quad m_j=\frac{v_x}{v_r},\quad k_j=\frac{\ell_x}{\ell_r},
\]
at $(z^{k+1},\theta_j)$. Let $t\downarrow u_x/u_r=m_j-c_x/v_r$, the type's indifference slope. Rearranging gives
\begin{equation}\label{eq:downward-iso-slope}
 k_j-m_j\leq\left(\frac{q_j}{\gamma_n}-1\right)
 \frac{c_x(x^{k+1},\theta_j;\tau_n)}{v_r(z^{k+1},\theta_j)}.
\end{equation}
On the compact set, $c_x\to0$ uniformly and $q_j,v_r,\gamma_n$ stay bounded and positive. The right side tends to zero, so the limiting upper bundle satisfies
\[
\frac{\ell_x(z^{k+1},\theta_j)}{\ell_r(z^{k+1},\theta_j)}
\leq\frac{v_x(z^{k+1},\theta_j)}{v_r(z^{k+1},\theta_j)}.
\]
This says that moving toward higher scope along type $j$'s iso-value curve weakly saves reviewer time. It is the slope condition needed to apply Assumption~\ref{ass:q} to the pair.

\emph{Step 5: Compare the two bundles' marginal verification productivity.}
We have the two ingredients for comparative verification advantage: downward indifference and the iso-value load slope. Since $c\to0$ on bounded scope, the binding IC equation gives
\[
v(z^k,\theta_j)-v(z^{k+1},\theta_j)\longrightarrow0.
\]
Thus the bundles share a limiting iso-value curve for $j$. If their scopes remain distinct, the slope restriction and the same-type condition in Assumption~\ref{ass:q} give $q(z^k,\theta_j)<q(z^{k+1},\theta_j)$. The pointwise condition $q_\theta>0$ then gives $q(z^k,\theta_i)<q(z^{k+1},\theta_h)$ for every lower-group type $i$ and upper-group type $h$. If the scope gap closes, $v_r>0$ forces the depth gap to close too, and $q_\theta>0$ gives the same strict ranking at the common bundle. In either case, positive load weights yield
\[
 Q_{B_k}<Q_{B_{k+1}}\qquad\text{in the limit}.
\]
Along the chain this gives $Q_{B_a}<Q_{B_b}$, contradicting the endpoint FOCs $Q_{B_a}\geq Q_{B_b}$ in Step 3. Compactness makes the contradiction uniform on $\mathcal T$; fixing $T$ gives the pointwise version.
\end{proof}

The full-information comparison in Theorem~\ref{thm:bottleneck} needs the frontier $V^{FI}$ to rise with technology; the benchmark in Section~\ref{sec:scarcity} also uses its eventual equality with $V^{FB}$. The next lemma establishes both by tracking how cheaper drafting relaxes participation.

\begin{lemma}[Full-information frontier]\label{lem:fi-frontier}
For every $T\geq0$, $V^{FI}(T,\tau)$ is nondecreasing in $\tau$; at
any fixed $T$ it equals $V^{FB}(T)$ for all sufficiently large $\tau$.
\end{lemma}

\begin{proof}
Observed quality removes IC. As $c$ falls, participation becomes easier, so $V^{FI}$ rises. For eventual equality with $V^{FB}$, take a bounded benchmark optimum and use $c(x,\theta;\tau)\to0$ on its finitely many positive bundles to make all participation constraints slack.

Assumption~\ref{ass:c} gives $c_{x\tau}<0$ and $c(0,\theta;\tau)=0$, hence $c_\tau(x,\theta;\tau)<0$ for $x>0$. Every allocation feasible under full information at $\tau_1$ remains feasible at $\tau_2>\tau_1$, giving $V^{FI}(T,\tau_2)\geq V^{FI}(T,\tau_1)$.

Choose an attained optimum for $V^{FB}(T)$, replacing zero-value entries by $(0,0)$. Load coercivity bounds its positive bundles. On their bounded scope, Assumption~\ref{ass:c} gives
\[
0\leq c(x,\theta;\tau)\leq x\sup_{0\leq s\leq x}c_x(s,\theta;\tau)\longrightarrow0.
\]
The same benchmark menu eventually satisfies IR and attains $V^{FB}(T)$ under full information, proving equality for all large $\tau$.
\end{proof}

\subsection{Proofs of the main results}

\begin{proof}[\textbf{Proof of Proposition~\ref{prop:microfoundation}}]\pagetarget{proof_prop:microfoundation}
The key issue is the same-type part of Assumption~\ref{ass:q}. We show that marginal verification productivity increases along an iso-value curve toward any broader assignment at which substituting scope for scrutiny weakly saves reviewer time.

\emph{Step 1: Check the basic verification properties and boundedness.}
Write $A(x,\theta)=a(\theta)p(x)$, $t=\lambda r$, and $g(t)=1-e^{-t}$. Define
\[
H(t)=\int_0^t(1-e^{-(t-s)})e^{-\alpha s}\,ds,
\qquad
K(t)=H'(t)=\int_0^t e^{-(t-s)}e^{-\alpha s}\,ds.
\]
Then
\[
v=Ag(t),\qquad
\ell_\alpha=\frac{x}{\lambda}[t-AH(t)],
\qquad
\ell_{\alpha,r}=x[1-AK(t)].
\]
Since $\alpha\geq0$, we have $0\leq H(t)\leq t-g(t)$ and $0<K(t)\leq g(t)<1$ for $t>0$. Thus $\ell_r>0$. The assumptions on $a$ and $p$ give $v_x,v_r>0$, bounded $v$, and $v=0$ on either axis.

For load coercivity, observe that
\[
\ell_\alpha\geq x\left[(1-A)r+\frac{Ag(t)}{\lambda}\right]
\geq\frac{xv}{\lambda}.
\]
Hence $v\geq\varepsilon>0$ and $\ell_\alpha\leq B$ imply $x\leq\lambda B/\varepsilon$. Since $p(0)=0$ and $v\leq A$, positive value also bounds $x$ away from zero. On the resulting bounded scope interval, $A$ is uniformly below one, so $\ell_\alpha\geq x(1-A)r$ bounds review depth. Positive-value, load-bounded bundles are therefore compact, establishing Assumption~\ref{ass:ell}.

\emph{Step 2: Check the verification-productivity ordering.}
The marginal product of verification satisfies
\[
q_\alpha(x,r,\theta)=\frac{\lambda A(x,\theta)e^{-t}}{x[1-A(x,\theta)K(t)]},
\qquad
(q_\alpha)_\theta=\frac{\lambda e^{-t}a'(\theta)p(x)}{x[1-AK(t)]^2}>0.
\]
Fix a type $\theta$ and a positive verified value $w$. Along the iso-value curve $A(x)g(t(x))=w$, write $A(x)=a(\theta)p(x)$, $E=e^{-t}$, and $B=1-AK(t)$. Then
\[
q_\alpha(x,r(x),\theta)=\frac{\lambda[A(x)-w]}{xB(x)},
\qquad
B'(x)=-\frac{A'(x)}{E}\bigl[K(t)-g(t)e^{-\alpha t}\bigr]\leq0,
\]
because $K(t)\geq g(t)e^{-\alpha t}$ follows from the integral defining $K$.

It remains to show that the reviewer-efficiency condition at a higher-scope bundle $x'$ makes $q_\alpha$ increase along the iso-value curve up to $x'$. Differentiating reviewer load along this curve, with $m=xA'/A$, gives
\[
\lambda\frac{d}{dx}\ell_\alpha(x,r(x),\theta)
=t-AH-m\bigl[AH+(e^t-1)B\bigr].
\]
At $m=E$, the right side equals
\[
t-g(t)-A\bigl[(1+E)H-g(t)K\bigr]
\geq(1-A)[t-g(t)]>0,
\]
where $g(t)K\geq EH$ and $H\leq t-g(t)$ follow from the integral representations. Since the reviewer-load derivative is strictly decreasing in $m$, the condition that greater scope locally saves reviewer time at $x'$ implies $m(x')>E(x')$, or
\[
x'A'(x')>A(x')-w.
\]
The curvature restriction on $p$ implies $A''<0$. Hence $xA'(x)-A(x)+w$ is strictly decreasing in $x$, so this inequality also holds for every $x<x'$ on the iso-value curve. Consequently,
\[
\frac{d}{dx}\log q_\alpha(x,r(x),\theta)
=\frac{A'(x)}{A(x)-w}-\frac1x-\frac{B'(x)}{B(x)}>0
\qquad (x<x').
\]
Thus $q_\alpha(z',\theta)>q_\alpha(z,\theta)$ for every narrower bundle $z$ on the same iso-value curve, as required by Assumption~\ref{ass:q}.

\emph{Step 3: Verify the researcher side.}
The specified drafting costs satisfy Assumption~\ref{ass:c}. The curvature restriction implies $p''<0$ for $x>0$, hence $p'(0)>0$. Since $\kappa'(0)=0$,
\[
\lim_{x\downarrow0}u_x
=a(\theta)p'(0)(1-e^{-\lambda r})>0
\]
for every $r>0$. We also have $u_\theta=a'(\theta)p(x)(1-e^{-\lambda r})>0$ and
\[
\frac{\partial}{\partial\theta}
\left(\frac{u_x}{u_r}\right)
=
\frac{e^{-\tau}\kappa'(x)a'(\theta)}
{a(\theta)^2p(x)\lambda e^{-\lambda r}}>0.
\]
Thus Assumption~\ref{ass:u} holds. Finally, $C(T)=T^2/2$ satisfies Assumption~\ref{ass:C}.
\end{proof}

\begin{proof}[\textbf{Proof of Proposition~\ref{prop:screening-distortions}}]\pagetarget{proof_prop:screening-distortions}
The two-type calculation in Section~\ref{sec:scarcity} gives the distortion signs when each type has its own bundle. With several types, the extreme assignments may be shared. We first sum the review-depth conditions for the extreme pooling groups to obtain the individual marginal-product inequalities. We then vary the strongest and weakest researchers' assignments \emph{one at a time}. Single crossing makes each variation relax the relevant pairwise incentive constraints, so the resulting first-order inequalities apply to the individual formats even under pooling.

\emph{Step 1: Order assignments and identify binding incentives.}
Fix $T>0$ and sufficiently advanced $\tau$. Lemmas~\ref{lem:uniform-optima} and~\ref{lem:downward} give interior assignments, slack participation, KKT multipliers, and slack downward IC between distinct bundles. By strict single crossing, $x_1\leq\cdots\leq x_M$; types assigned the same bundle form consecutive pooling groups. Every binding cross-group IC therefore runs upward, from a lower type toward a higher-scope assignment. Write $\lambda_{ih}\geq0$ for type $i$'s incentive constraint against entry $h$ and $\gamma\geq0$ for the reviewer-capacity multiplier.

\emph{Step 2: Establish the individual scrutiny inequalities.}
For a pooling group $B$ with common bundle $z^B$, define
\[
 W_{B,r}=\sum_{i\in B}f_i v_r(z^B,\theta_i),\qquad
 L_{B,r}=\sum_{i\in B}f_i\ell_r(z^B,\theta_i),\qquad
 Q_B=\frac{W_{B,r}}{L_{B,r}}.
\]
Adding the review-depth first-order conditions for all entries in $B$ cancels multipliers on IC between types sharing that assignment. The remaining conditions involve only imitation across groups:
\begin{equation}\label{eq:screening-block-stationarity}
 W_{B,r}-\gamma L_{B,r}
 =-\sum_{\substack{i\in B\\h\notin B}}\lambda_{ih}u_r(z^B,\theta_i;\tau)
 +\sum_{\substack{i\notin B\\h\in B}}\lambda_{ih}u_r(z^B,\theta_i;\tau).
\end{equation}
The lowest group $B_L$ can have outgoing upward IC multipliers, while the highest group $B_H$ can have incoming ones. Since $u_r=v_r>0$ and $L_{B,r}>0$, equation~\eqref{eq:screening-block-stationarity} implies
\[
 Q_{B_L}\leq\gamma\leq Q_{B_H},\qquad \gamma>0.
\]
Assumption~\ref{ass:q} gives $q_\theta>0$ at a common bundle. Each $Q_B$ is a positive-weighted average of the individual marginal products $q=v_r/\ell_r$, so
\begin{equation}\label{eq:screening-individual-q}
 q(z_1,\theta_1)\leq Q_{B_L}\leq\gamma
 \leq Q_{B_H}\leq q(z_M,\theta_M).
\end{equation}
Thus the weakest type is over-scrutinized and the strongest under-scrutinized relative to $q=\gamma$, including when either shares an assignment with other types.

\emph{Step 3: Separate the strongest type by a feasible one-sided change.}
At the strongest type's assigned bundle $z_M$, strict single crossing gives an interval
\[
 \frac{u_x(z_M,\theta_{M-1};\tau)}{u_r(z_M,\theta_{M-1};\tau)}
 <t<
 \frac{u_x(z_M,\theta_M;\tau)}{u_r(z_M,\theta_M;\tau)}.
\]
Increase only type $M$'s scope by $h>0$ and reduce her review depth by $th$. The strongest type gains private utility at rate $u_r(z_M,\theta_M;\tau)(u_x/u_r-t)>0$. Every lower type loses utility from imitating that modified entry, since its $u_x/u_r$ is below $t$. Thus all pairwise IC constraints remain satisfied for sufficiently small $h$, whether $z_M$ was initially pooled or distinct. Participation is slack.

The editor's Lagrangian for~\eqref{eq:editor-problem} has a zero derivative along the chosen direction. Because the perturbation weakly relaxes every active IC constraint, its derivative of verified value \emph{net of reviewer time priced at $\gamma$} is nonpositive:
\[
 v_x(z_M,\theta_M)-t v_r(z_M,\theta_M)
 -\gamma\bigl[\ell_x(z_M,\theta_M)-t\ell_r(z_M,\theta_M)\bigr]\leq0.
\]
Let $t$ decrease to the lower endpoint of the interval and rearrange, using the iso-value slope $dr/dx=-v_x/v_r$ from~\eqref{eq:efficient-format}:
\begin{align}\label{eq:screening-individual-top-format}
 &\gamma\left[\ell_x(z_M,\theta_M)
 -\frac{v_x(z_M,\theta_M)}{v_r(z_M,\theta_M)}\ell_r(z_M,\theta_M)\right]\\
 &\quad\geq
 \left[\frac{v_x(z_M,\theta_M)}{v_r(z_M,\theta_M)}
 -\frac{u_x(z_M,\theta_{M-1};\tau)}{u_r(z_M,\theta_{M-1};\tau)}\right]
 \left[v_r(z_M,\theta_M)-\gamma\ell_r(z_M,\theta_M)\right]\geq0.\notag
\end{align}
The first bracket on the right is positive: $v_x/v_r\geq u_x/u_r$ for the strongest type because $c_x\geq0$, and $u_x/u_r$ rises strictly with type by Assumption~\ref{ass:u}. The second bracket is nonnegative by~\eqref{eq:screening-individual-q}. The expression on the left is the strongest \emph{individual} type's marginal reviewer load along her own iso-verified-value curve. Its nonnegative sign establishes a scope-intensive format. It is strictly positive whenever $q(z_M,\theta_M)>\gamma$.

\emph{Step 4: Separate the weakest type under separability.}
Now impose~\eqref{eq:separable}. At the weakest type's bundle $z_1$, select
\[
 \frac{u_x(z_1,\theta_1;\tau)}{u_r(z_1,\theta_1;\tau)}
 <t<
 \frac{u_x(z_1,\theta_2;\tau)}{u_r(z_1,\theta_2;\tau)}.
\]
Reduce only type 1's scope by $h>0$ and increase review depth by $th$. Type 1 gains utility, whereas all higher types find the modified bundle less attractive. The change therefore relaxes the active pairwise IC constraints, including within a pooled lowest group. The corresponding net-value first-order inequality is
\[
 -\bigl[v_x(z_1,\theta_1)-\gamma\ell_x(z_1,\theta_1)\bigr]
 +t\bigl[v_r(z_1,\theta_1)-\gamma\ell_r(z_1,\theta_1)\bigr]\leq0.
\]
Let $t$ increase to the upper endpoint and rearrange:
\begin{align}\label{eq:screening-individual-bottom-format}
 &\gamma\left[\ell_x(z_1,\theta_1)
 -\frac{v_x(z_1,\theta_1)}{v_r(z_1,\theta_1)}\ell_r(z_1,\theta_1)\right]\\
 &\quad\leq
 \left[\frac{v_x(z_1,\theta_1)}{v_r(z_1,\theta_1)}
 -\frac{u_x(z_1,\theta_2;\tau)}{u_r(z_1,\theta_2;\tau)}\right]
 \left[v_r(z_1,\theta_1)-\gamma\ell_r(z_1,\theta_1)\right]\leq0.\notag
\end{align}
Under separability, $v_x/v_r=y_x/y_r$ for every type, whereas
$u_x(z_1,\theta_2;\tau)/u_r(z_1,\theta_2;\tau)
=y_x/y_r-c_x(x_1,\theta_2;\tau)/(a_2y_r)\leq y_x/y_r$.
The first bracket on the right is therefore nonnegative; the second is nonpositive by~\eqref{eq:screening-individual-q}. The weakest \emph{individual} type's iso-value reviewer-load slope is nonpositive, establishing a review-heavy format. All four inequalities in the proposition thus hold for the extreme types themselves, even when their assignments are pooled.
\end{proof}

\begin{proof}[\textbf{Proof of Theorem~\ref{thm:cost-dependent-separation}}]\pagetarget{proof_thm:cost-dependent-separation}
Fixing both an imitation advantage and positive verified value forces reviewer load above every finite budget. The key issue is bounding the high bundle's scope under a supposed finite budget. Load coercivity provides that bound; pairwise IC then contradicts $c_x\to0$ on bounded scope sets.

Suppose menus at $\tau_n\to\infty$ have load at most $B$ and satisfy the theorem's thresholds $\delta,\varepsilon>0$. The high bundle has $v(z_j^n,\theta_j)\geq\varepsilon$ and $\ell(z_j^n,\theta_j)\leq B/f_j$. Assumption~\ref{ass:ell} bounds inputs whenever value is positive and load bounded, so $x_j^n\leq X<\infty$.

The weak type's pairwise IC requires her gross imitation gain $\delta$ to be covered by additional drafting cost. But $c_x\to0$ uniformly on $[0,X]$ by Assumption~\ref{ass:c}, so
\[
 \delta\leq c(x_j^n,\theta_i;\tau_n)-c(x_i^n,\theta_i;\tau_n)
 \leq c(x_j^n,\theta_i;\tau_n)
 \leq X\sup_{\theta\in\Theta,\,x\in[0,X]}c_x(x,\theta;\tau_n).
\]
The right side tends to zero, contrary to $\delta>0$. Thus every sufficiently advanced qualifying menu exceeds the chosen budget $B$.
\end{proof}

\begin{proof}[\textbf{Proof of Theorem~\ref{thm:fixed-capacity}}]\pagetarget{proof_thm:fixed-capacity}
We compare \emph{optimal menus}, allowing the editor to redesign every assignment. At large $\tau$, participation and downward cross-bundle IC are slack (the two preceding lemmas), while upward IC tightens as $c_x$ falls. A later optimal menu is therefore feasible at slightly earlier technology. Upper semicontinuity extends that local comparison to arbitrary advanced technology pairs.

Let $\mathcal T=[T_-,T_+]$ be given by Lemma~\ref{lem:uniform-optima} and $\tau_{\mathrm{IC}}$ by Lemma~\ref{lem:downward}.
 
\emph{Step 1: Compare a later optimal menu with an earlier technology.}
Fix $T\in\mathcal T$ and $\tau>\tau_{\mathrm{IC}}$.
At the optimum, IR and downward cross-bundle IC are slack. Every binding external IC runs upward: a weak type is deterred by the cost of the higher scope. Since $c_{x\tau}<0$, lowering $\tau$ slightly increases that cost difference and relaxes upward IC. Slack constraints stay slack by continuity; within-pool IC remains equality.

The same menu thus works at all $s<\tau$ sufficiently close, with identical value and load. Optimality at $s$ gives
\[
V^*(T,s)\geq V^*(T,\tau)\qquad (s<\tau\text{ sufficiently close}).
\tag{A1}
\] \emph{Step 2: Extend the local comparison to any two advanced technologies.}
The neighborhood in (A1) depends on $\tau$. Compactness of optimal menus and continuity of the constraints make $g(\tau)=V^*(T,\tau)$ upper semicontinuous. An upward move in $g$ would have a first crossing of some intermediate value, which (A1) rules out.

For precision, write $g(\tau)=V^*(T,\tau)$ and suppose $g(b)>g(a)$ at two advanced technologies $a<b$. Choose a level $\alpha$ between them. Upper semicontinuity makes the superlevel set $\{t\in[a,b]:g(t)\geq\alpha\}$ closed, so it has a first point $t_*>a$. All earlier points have values below $\alpha$, yet (A1) says that immediately to the left of $t_*$ the value must be at least $g(t_*)\geq\alpha$. This contradiction establishes global weak decline.
Writing $\bar\tau:=\tau_{\mathrm{IC}}$
(and increasing it if necessary so that Lemma~\ref{lem:uniform-optima} and the constraint qualification in Lemma~\ref{lem:downward} apply), we have
\[
V^*(T,\tau_2)\leq V^*(T,\tau_1)
\qquad
(T\in\mathcal T,\ \tau_2>\tau_1>\bar\tau).
\tag{A2}
\]
\end{proof}

\begin{proof}[\textbf{Proof of Theorem~\ref{thm:ci-equalization}}]\pagetarget{proof_thm:ci-equalization}
Under separability, pairwise IC bounds the success difference $y_j-y_i$ by normalized drafting-cost differences. Compactness bounds the scope and Assumption~\ref{ass:c} makes their incremental costs vanish. The key issue is applying these bounds uniformly over \emph{all} optimal menus to obtain a common limit.

Theorem~\ref{thm:fixed-capacity} gives eventually decreasing $V^*(T,\tau)$, and the pooled menu in Lemma~\ref{lem:uniform-optima} gives a positive lower bound. Therefore $V^*(T,\tau)\to V_T>0$.

Under $v=a_i y$, utility is $u_i=a_i y-c_i(x;\tau)$. The two pairwise IC inequalities~\eqref{eq:ci-ic-interval} bound $y_j-y_i$ between the incremental drafting costs $[c_j(x_j)-c_j(x_i)]/a_j$ and $[c_i(x_j)-c_i(x_i)]/a_i$. Single crossing gives $x_j\geq x_i$; compactness bounds both scope. Since $c_x\to0$ uniformly on this bounded set, both endpoints tend to zero, hence $y_j-y_i\to0$ for every $i<j$.

Since $V^*=\sum_i f_i a_i y_i\to V_T$, the common success limit is $y_T=V_T/\sum_i f_i a_i$. Uniform bounds over optimal menus give the conclusion for every optimal selection.
\end{proof}

\begin{proof}[\textbf{Proof of Theorem~\ref{thm:ci-pooling}}]\pagetarget{proof_thm:ci-pooling}
Equal success permits distinct research--review formats. We first characterize the limiting formats as a minimum-load allocation subject to scope sorting. The key issue for the pooling case is stronger: convergence to a common format must imply \emph{exact} pooling at large finite $\tau$. Closing a residual scope gap yields a first-order load saving that dominates its $O(c_x)$ screening benefit.

\emph{Step 1: Write pairwise IC in terms of success gaps.}
Let $a_i=a(\theta_i)$, $A=\sum_i f_i a_i$, and for a positive attainable success level $y$ let $r(x,y)$ be the review depth solving $y(x,r)=y$. For the proof alone write
\[
\mathcal L_i(x,y)=\ell(x, r(x,y), \theta_i).
\]
The weak type's upward IC places an upper bound on $y_H-y_L$; the strong type's downward IC places a lower bound. Together,
\begin{equation}\label{eq:ci-proof-increments}
\frac{c(x_H,\theta_H;\tau)-c(x_L,\theta_H;\tau)}{a_H}
\leq y_H-y_L\leq
\frac{c(x_H,\theta_L;\tau)-c(x_L,\theta_L;\tau)}{a_L}.
\end{equation}
Single crossing under separability implies $c_{Hx}/a_H<c_{Lx}/a_L$, so the IC interval is nonempty. To quantify how it shrinks on bounded scope, set
\[
\varepsilon_\tau=\max_{i=L,H}\sup_{0\leq x\leq X}
\frac{c_x(x,\theta_i;\tau)}{a_i}\longrightarrow0.
\]
Then $0\leq y_H-y_L\leq\varepsilon_\tau(x_H-x_L)$.

\emph{Step 2: Identify the limiting minimum-load allocation.}
Let $R_\tau(V)$ be the minimum load needed to deliver exact aggregate value $V$ under IC and IR. Theorem~\ref{thm:ci-equalization} and $x_L\leq x_H$ suggest the limiting problem: deliver common success $V/A$ with the least load, respecting scope order,
\begin{equation}\label{eq:ci-ordered-problem}
P(V)=\min_{x_L\leq x_H}
\left\{f_L\mathcal L_L(x_L,V/A)+f_H\mathcal L_H(x_H,V/A)\right\}.
\end{equation}
To justify this limit, fix $\tau_n\to\infty$ and targets $V_n\to V_0>0$. Start from any ordered scope pair $(\widehat x_L,\widehat x_H)$. Choose a success gap in the pairwise IC interval~\eqref{eq:ci-proof-increments}, then shift both success levels by a common constant so $\sum_i f_i a_i y_i=V_n$. The gap is $O(\varepsilon_{\tau_n})$, so $y_i\to V_0/A$. Since $y_r>0$, nearby review depths implement these success indices; IR holds for large $n$ because $c\to0$. Thus every ordered format has feasible finite-technology approximations. Applying this construction to the least-load format gives
\[
\limsup_{n\to\infty}R_{\tau_n}(V_n)\leq P(V_0).
\]
For the converse, these comparison menus bound the loads of actual $R_{\tau_n}(V_n)$ minimizers. Because $V_n\to V_0>0$, some bundle has positive value. Pairwise IC transfers a positive payoff to the other type once $c\to0$. Assumption~\ref{ass:ell} then bounds both bundles in a compact interior set. Every convergent subsequence has ordered scope and common success $V_0/A$ by Step 1, so its load is at least $P(V_0)$. Together with the upper bound,
\[
 R_{\tau_n}(V_n)\longrightarrow P(V_0),
\]
Every limit of minimum-load menus solves the ordered problem. Compactness also gives attainment of the minimum at the relevant targets.

\emph{Step 3: Link the minimum-load problem back to the editor.}
Now link this fixed-target problem to the editor's optimum $V_\tau=V^*(T,\tau)$. Under separability, adding the same small $dy>0$ to every success index increases each type's utility from \emph{every} menu option by $a_i dy$. Pairwise IC is unchanged, while aggregate value rises by $A dy$. Thus an interior editorial optimum exhausts capacity. A menu delivering $V_\tau$ with less load would permit such an increase, so every editorial optimum minimizes load at its value:
\[
 R_\tau(V_\tau)=T.
\]
As $V_\tau\to Ay_T$ by Theorem~\ref{thm:ci-equalization}, Step 2 applies also to these moving targets. Optimal assignments therefore approach the ordered minimizer at $Ay_T$.

\emph{Step 4: Determine the limiting ordering with two types.}
At success $y_T$, strict convexity makes each type's load-minimizing scope $x_i^E(y_T)$ unique. If $x_L^E<x_H^E$, those preferred formats already respect $x_L\leq x_H$; the ordered problem chooses both, hence
\[
 x_H^*(T,\tau)-x_L^*(T,\tau)
 \longrightarrow x_H^E(y_T)-x_L^E(y_T)>0.
\]
If $x_L^E>x_H^E$, the preferred formats reverse the IC scope order. Strict convexity puts the ordered minimizer on the binding boundary $x_L=x_H=x^P$, between the two individual optima. Therefore
\[
 x_H^E(y_T)<x^P<x_L^E(y_T),\qquad
 \mathcal L_{H,x}(x^P,y_T)>0.
\]
Step 3 gives convergence toward $(x^P,r(x^P,y_T))$. We next close any residual finite-technology gap.

\emph{Step 5: Prove exact finite-technology pooling.}
Suppose an advanced optimum has scope gap $h=x_H-x_L>0$. Give the strong type scope $x_L$ and set both success indices to $V_\tau/A$. The new bundles coincide, so pairwise IC holds; aggregate verified value stays exactly $V_\tau$, and IR remains slack. We compare their reviewer loads.

At the pooled limit, $\mathcal L_{H,x}(x^P,y_T)>0$, so removing $h$ from the strong type's scope saves at least $f_H\delta h$ in reviewer load locally. The success adjustment costs at most $C\varepsilon_\tau h$: Step 1 bounds $y_H-y_L\leq\varepsilon_\tau h$, and keeping $\sum_i f_i a_i y_i$ fixed makes each index adjustment of the same order. Thus
\[
\begin{aligned}
\Delta L
&=f_H\bigl[\mathcal L_H(x_L,V_\tau/A)-\mathcal L_H(x_H,y_H)\bigr]
+f_L\bigl[\mathcal L_L(x_L,V_\tau/A)-\mathcal L_L(x_L,y_L)\bigr]\\
&\leq -f_H\delta h+C\varepsilon_\tau h
\end{aligned}
\]
Here $\delta>0$ bounds the marginal scope saving below, and $C<\infty$ bounds the marginal load cost of changing success. Since $\varepsilon_\tau\to0$, the saving dominates: $\Delta L<0$. This contradicts $R_\tau(V_\tau)=T$ in Step 3. Thus $x_H=x_L$; pairwise IC gives $y_H=y_L$, and $y_r>0$ gives $r_H=r_L$.
\end{proof}

\begin{proof}[\textbf{Proof of Proposition~\ref{prop:existence}}]\pagetarget{proof_prop:existence}
Equilibrium requires an attained editorial value $V^*(T,\tau)$ and a positive zero of the reviewer-supply residual $F=C_T-\phi V^*/T$. The key issue is that editorial choices are unbounded and $V^*$ may jump as capacity varies. We compactify feasible menus, establish the sign of $F$ near zero and infinity, and use its one-sided continuity to find the greatest root.

\pagetarget{proof:existence}{}
\emph{Step 1: Ensure that the editor's optimum exists and varies appropriately with capacity.}
Fix finite $\tau$ and a capacity bound $T_0$. IR gives $c(x_i,\theta_i;\tau)\leq v(x_i, r_i, \theta_i)\leq\sup v$. Since drafting cost is coercive in $x$ (Assumption~\ref{ass:c}), every feasible menu has bounded scope. Review depth can still diverge when scope tends to zero, so we treat such entries separately.
 
Take a load-bounded maximizing sequence. Along a subsequence, each $x_i^n$ converges and each $r_i^n$ converges or tends to infinity. If $r_i^n\to\infty$, Assumption~\ref{ass:ell} (positive value plus bounded load bounds the inputs) forces $v(z_i^n,\theta_i)\to0$. As $v$ increases in both inputs, this also forces $x_i^n\to0$. Thus $u(z_i^n,\theta_i;\tau)\to0$, while any other type's utility from that entry has nonnegative lower limit, since $v\geq0$ and $c(0,\theta;\tau)=0$. Replace each escaping entry by $(0,0)$ in the limiting menu. A type whose entry escaped already weakly preferred it to every alternative, so the limiting null entry preserves its IC. For other types, the null alternative has utility zero and their participation constraints ensure they prefer their own entries. Other IC constraints pass to the limit by continuity. Discarding the escaping entries' loads preserves capacity, and their verified values tend to zero. The limiting menu attains the maximizing sequence's value.
 
This gives an attained optimum. Since capacity enlarges the feasible set, $V^*(T,\tau)$ is nondecreasing in $T$; the same compactification applied to $T_n\downarrow T$ gives right-continuity.
 
\emph{Step 2: Locate the reviewer-supply residual below and above zero.}
The small pooled menu from Lemma~\ref{lem:uniform-optima} supplies a lower bound on impact. For fixed $r_0>0$, Assumption~\ref{ass:u} gives $u_x>0$ as $x\downarrow0$, hence pooled verified value $\geq ax$ for small $x$. Assumption~\ref{ass:ell} gives $\ell(0, r_0, \theta)=0$ and finite $\ell_x$, hence pooled load $\leq bx$. With $x=T/b$, we get $V^*(T,\tau)/T\geq a/b>0$ near $T=0$.

Since $C_T(0)=0$, the residual $F=C_T-\phi V^*/T$ is negative near zero. As $T\to\infty$, $V^*\leq\bar V$ while $C_T(T)\to\infty$, so $F>0$ eventually. The remaining task is to show that a change from negative to positive crosses zero.
 
\emph{Step 3: Prove an equilibrium crossing and a largest crossing.}
Right-continuity of $V^*$ makes $F(\cdot;\tau)$ right-continuous. Moreover, for $T_n\uparrow T$, monotonicity of $V^*$ gives
\begin{equation}\label{eq:no-upward-jumps}
 F(T;\tau)\leq\liminf_{n\to\infty}F(T_n;\tau).
\end{equation}
Graphically, $F$ may jump downward as capacity unlocks a better menu, but inequality~\eqref{eq:no-upward-jumps} precludes upward jumps. Let $t_0=\inf\{T:F(T;\tau)>0\}$. The sign bounds make $t_0$ finite and positive. Right-continuity excludes $F(t_0)<0$, and the no-upward-jump inequality excludes $F(t_0)>0$. Thus $F(t_0)=0$.
 
Zeros are bounded above since $F>0$ eventually. Let $b_0$ be their supremum. If zeros approach from below,~\eqref{eq:no-upward-jumps} gives $F(b_0)\leq0$. Were it negative, eventual positivity would yield a further zero above $b_0$, a contradiction. Thus $F(b_0)=0$; if the zeros have a largest isolated member, that member is $b_0$.
\end{proof}

\begin{proof}[\textbf{Proof of Theorem~\ref{thm:bottleneck}}]\pagetarget{proof_thm:bottleneck}
Theorem~\ref{thm:fixed-capacity} shifts the reviewer-supply residual $F=C_T-\phi V^*/T$ upward at each fixed capacity. The key issue is equilibrium multiplicity: we compare the \emph{greatest} zero using the crossing property from Proposition~\ref{prop:existence}, then use $V=TC_T/\phi$ to compare verified knowledge.

\emph{Step 1: Compare private-information residuals and their greatest zeros.} 
For $\tau_2>\tau_1>\bar\tau$ and $T\in\mathcal T$, Theorem~\ref{thm:fixed-capacity} gives $V^*(T,\tau_2)\leq V^*(T,\tau_1)$, hence $F(T;\tau_2)\geq F(T;\tau_1)$. The crossing argument gives $F(T;\tau)>0$ above the greatest zero: any subsequent nonpositive value would force another zero farther right.
 
If $\overline T(\tau_2)>\overline T(\tau_1)$, both capacities belong to $\mathcal T$ and $0=F(\overline T(\tau_2);\tau_2) \geq F(\overline T(\tau_2);\tau_1)>0$, a contradiction. Hence $\overline T(\tau_2)\leq\overline T(\tau_1)$. The equilibrium identity $V=TC_T(T)/\phi$ and the strict positivity of $C_T+TC_{TT}$ establish the verified-value inequality.

\emph{Step 2: Reverse the comparison when quality is observed.}
The compactification and crossing argument of Proposition~\ref{prop:existence} also applies under full information: $V^{FI}(\cdot,\tau)$ is attained, nondecreasing and right-continuous, and its supply residual has a greatest positive zero. The same holds for $V^{FB}$.
 
Lemma~\ref{lem:fi-frontier} gives $V^{FI}(T,\tau_2)\geq V^{FI}(T,\tau_1)$: the full-information residual shifts downward, so its greatest zero rises, $\overline T^{FI}(\tau_2)\geq\overline T^{FI}(\tau_1)$. Also $V^{FI}\geq V^*$, hence $\overline T^{FI}(\tau)\geq\overline T(\tau)$. Because $TC_T(T)/\phi$ increases strictly, both comparisons carry over to equilibrium verified value.
 
Let $\overline T^{FB}$ be the greatest equilibrium using the technology-independent frontier $V^{FB}$. Since $V^{FI}\leq V^{FB}$, we have $\overline T^{FI}(\tau)\leq\overline T^{FB}$. Lemma~\ref{lem:fi-frontier} gives $V^{FI}(\overline T^{FB},\tau)=V^{FB}(\overline T^{FB})$ for large $\tau$, making $\overline T^{FB}$ a full-information equilibrium then. Hence $\overline T^{FI}=\overline T^{FB}$ and the corresponding equilibrium verified value is eventually constant.
\end{proof}

\begin{proof}[\textbf{Proof of Theorem~\ref{thm:strict-M}}]\pagetarget{proof_thm:strict-M}
Strict decline requires a positive screening cost at every sufficiently advanced optimum. We first use the type ordering of iso-success format costs to rule out full pooling. Next we show that scrutiny's marginal productivity differs across distinct assignments; summing FOCs within pooling groups then forces a positive upward-IC multiplier. Earlier technology relaxes that IC, permitting a profitable reallocation of review. The equilibrium result follows from the supply residual.

\emph{Step 1: Translate the primitive condition into a type ordering.}
Write
\[
D_i(x,r)=\ell_x(x, r, \theta_i)
 -\ell_r(x, r, \theta_i)\frac{y_x(x,r)}{y_r(x,r)}.
\]
The condition in Theorem~\ref{thm:strict-M} implies
\begin{equation}\label{eq:strict-format-advantage}
\partial_\theta D(x, r, \theta)
=\ell_{x\theta}-\ell_{r\theta}\frac{y_x}{y_r}<0,
\qquad D_{i+1}(x,r)<D_i(x,r).
\end{equation}
Here $D_i$ is the reviewer-load derivative along an iso-success curve, $dr/dx=-y_x/y_r$. As $\ell_{r\theta}<0$, multiplying the theorem's ratio inequality by $\ell_{r\theta}$ reverses its sign, yielding $D_\theta<0$. Higher types therefore have a stronger reviewer-time advantage in scope-intensive formats.

\emph{Step 2: Give the highest type a profitable pairwise departure from pooling.}
Suppose all types pool at a positive bundle $(x,r)$ with common success $y$. As in the pooling theorem, the editor exhausts capacity and minimizes load at attained value $V=Ay$, $A=\sum_i f_i a_i$. Along a common iso-success movement, an interior pooled optimum satisfies
\begin{equation}\label{eq:pool-foc}
  \sum_i f_i D_i(x,r)=0.
\end{equation}
By $D_{i+1}<D_i$ and the zero weighted average in~\eqref{eq:pool-foc}, $D_M(x,r)<0$: shifting the strongest type toward more scope and less review saves time. Uniform interior compactness and strict $D_\theta<0$ give a common margin, $f_MD_M\le-\delta<0$, across advanced pooled optima.

Give type $M$ scope $x+h$ and leave lower types at $x$. Pairwise IC across the new boundary requires a small success premium $d_h$ for type $M$. With $c_i(x;\tau)=c(x,\theta_i;\tau)$, the IC interval is
\[
\frac{c_M(x+h;\tau)-c_M(x;\tau)}{a_M}
\ \le d_h\le\
\frac{c_{M-1}(x+h;\tau)-c_{M-1}(x;\tau)}{a_{M-1}}.
\]
Single crossing gives $c_{Mx}/a_M<c_{M-1,x}/a_{M-1}$, so the interval is nonempty; vanishing marginal drafting costs give $0\le d_h\le\varepsilon_\tau h$ with $\varepsilon_\tau\to0$. To preserve aggregate verified value, set $\alpha=f_Ma_M/A$ and assign success $y-\alpha d_h$ to lower types and $y+(1-\alpha)d_h$ to type $M$. Their weighted average remains $y$, while the success gap $d_h$ satisfies IC. Review depths adjust locally because $y_r>0$, and IR stays slack.

The change in reviewer load is
\[
\Delta L=f_MD_M(x,r)h+O(\varepsilon_\tau h)+o(h),
\]
Since $f_MD_M\le-\delta$, the $h$ term is a first-order load saving; $O(\varepsilon_\tau h)$ is the IC adjustment cost and $o(h)$ the Taylor remainder. Choose large $\tau$ and then small $h>0$ to get $\Delta L<0$. This contradicts the pooled menu's minimum-load property, uniformly over $T\in\mathcal T$. Every advanced optimum therefore has at least two distinct assignments.

\emph{Step 3: Compare adjacent distinct assignments before aggregating.}
We seek $Q_B<Q_C$ for adjacent distinct assignments $B<C$, where $Q_B$ is the group's marginal product of review. Suppose $Q_B\ge Q_C$ persists along $\tau_n\to\infty$. By finite types and uniform compactness, pass to a subsequence with fixed adjacent groups and converging bundles. The IC success bounds give a common limiting success $y_0=V_0/A>0$.

As in Theorem~\ref{thm:ci-pooling}, the limit of value-maximizing menus minimizes reviewer load at its common success level $y_0$. Here we need the finite-type version of that argument. Begin with any ordered scope vector and choose small success increments within the normalized drafting-cost intervals for each \emph{adjacent} pair. Center those increments so aggregate value stays fixed and recover review depths using $y_r>0$. Single crossing orders $c_x/a_i$, so adding adjacent inequalities also satisfies incentive constraints between distant types. This constructs finite-technology menus approaching any candidate ordered format. Conversely, compactness, pairwise IC, and vanishing drafting costs force each limiting optimal menu to have common success $y_0$ and ordered scope. The two comparisons identify the ordered reviewer-time problem
\[
P(V_0)=\min_{x_1\le\cdots\le x_M}\sum_i f_i\mathcal L_i(x_i,V_0/A),
\]
and shows that the limiting editor-optimal menu solves it.

There are two possible limiting geometries for adjacent groups. Their scope gap may shrink to zero. In that case their review depths also coincide because both groups have the same success index and $y_r>0$. The assumption $q_\theta>0$ directly gives higher marginal reviewer productivity for every higher type at this common bundle: $q(z^B,\theta_i)<q(z^C,\theta_j)$ for $i\in B$, $j\in C$.

Alternatively, the gap remains positive. Let $j$ be the lowest type in the upper group. In the limiting ordered reviewer-time problem, a small reduction in $j$'s scope alone respects sorting: the preceding group has strictly lower scope, while the types above $j$ remain at least as scope-intensive. Optimality requires this feasible reduction to weakly increase reviewer time. Therefore the marginal load from \emph{increasing} $j$'s scope along her iso-success curve is weakly negative, $\mathcal L_{j,x}(x^C,y_0)\le0$. This is the slope inequality
\[
\frac{\ell_x(z^C,\theta_j)}{\ell_r(z^C,\theta_j)}
\le\frac{v_x(z^C,\theta_j)}{v_r(z^C,\theta_j)}.
\]
Both bundles deliver the same success index, so for type $j$ they also deliver the same verified value. The same-type condition in Assumption~\ref{ass:q}, together with the reviewer-efficiency inequality at the higher-scope bundle, gives $q(z^B,\theta_j)<q(z^C,\theta_j)$. The pointwise condition $q_\theta>0$ then gives $q(z^B,\theta_i)<q(z^C,\theta_h)$ for every lower-group type $i$ and upper-group type $h$. In either case the limiting productivity of every type in $C$ strictly exceeds that of every type in $B$. Weighted averaging yields $Q_B<Q_C$ in the limit, contrary to the weak reverse inequality along the sequence. The contradiction establishes a common finite threshold.

\emph{Step 4: Show that some pairwise incentive constraint is costly.}
Increase the technology threshold so that the two conclusions just proved, the uniform regularity conclusions, and Theorem~\ref{thm:fixed-capacity} hold. Then every optimal menu at any $T\in[T_-,T_+]$ has at least two distinct consecutive blocks with strictly ordered marginal verification productivities.

Suppose all cross-group IC multipliers are zero. Sum the review FOCs within each pooling group; internal multipliers cancel and IR is slack. The resulting stationarity condition is
\[
  \sum_{i\in B}f_i\bigl[v_r(z^B,\theta_i)-\gamma\ell_r(z^B,\theta_i)\bigr]=0,
  \qquad\text{equivalently}\qquad Q_B=\gamma,
\]
where $\gamma$ prices capacity. Step 3 gives $Q_B<Q_C$ across adjacent groups, contradicting $Q_B=\gamma$ for every group. Hence some cross-group IC multiplier is positive; Lemma~\ref{lem:downward} makes it upward IC. To identify the appropriate donor and recipient of reviewer time, choose $B_-$ as the \emph{lowest group with a positive outgoing upward multiplier}, and $B_+$ as the \emph{highest group with a positive incoming upward multiplier}. These groups need not be the two ends of the same pairwise constraint. Since $B_-$ has no positive incoming multiplier from a lower group, its summed review-depth FOC gives
\[
 W_{B_-,r}-\gamma L_{B_-,r}<0,\qquad Q_{B_-}<\gamma.
\]
Since $B_+$ has no positive outgoing multiplier to a higher group, its FOC gives
\[
 W_{B_+,r}-\gamma L_{B_+,r}>0,\qquad Q_{B_+}>\gamma.
\]
Thus $Q_{B_-}<\gamma<Q_{B_+}$. Move to slightly earlier $s<\tau_2$: the higher-scope drafting-cost differences rise, creating strict slack in every previously binding cross-group upward IC constraint. Transfer a small load $dL$ from $B_-$ to $B_+$, adjusting their common review depths and keeping total load fixed. All between-group IC remains slack for a sufficiently small transfer, and within-group IC remains satisfied because each group moves together. Verified value rises by $(Q_{B_+}-Q_{B_-})dL+o(dL)>0$, hence $V^*(T,s)>V^*(T,\tau_2)$. The weak monotonicity of Theorem~\ref{thm:fixed-capacity} extends this strict comparison to every $\tau_1\in(\tau_s,\tau_2)$:
\[
  V^*(T,\tau_2)<V^*(T,\tau_1).
\]

\emph{Step 5: Pass from a strict frontier loss to a strict equilibrium loss.}
Let $T_2=\overline T(\tau_2)$. At this capacity the earlier-technology supply residual is strictly negative:
\[
 C_T(T_2)-\phi\frac{V^*(T_2,\tau_1)}{T_2}
 <C_T(T_2)-\phi\frac{V^*(T_2,\tau_2)}{T_2}=0.
\]
The residual-crossing argument then gives an equilibrium at $\tau_1$ above $T_2$, so $\overline T(\tau_1)>\overline T(\tau_2)$. Since equilibrium verified knowledge is $TC_T(T)/\phi$, strictly increasing in $T$, verified knowledge also falls strictly.
\end{proof}

\begin{proof}[\textbf{Proof of Proposition~\ref{prop:institutional-improvements}}]\pagetarget{proof_prop:institutional-improvements}
Both improvements expand the editor's feasible menu set at fixed reviewer capacity. The key issue is translating the pointwise frontier gain into a comparison of greatest equilibrium capacities, using the same residual-crossing argument as Theorem~\ref{thm:bottleneck}.

At fixed $T,\tau$, verification improvement means $\ell_{r\eta}<0$ and $\ell(x, 0, \theta;\eta)=0$ in~\eqref{eq:verification-improvement}. Integrating in $r$ gives $\ell(x, r, \theta;\eta_2)\leq\ell(x,\theta,r;\eta_1)$: every old menu fits the new capacity constraint. A finer preliminary signal imposes IC only within smaller type cells, so it also preserves all previously feasible menus. Thus $V^*(T)$ rises weakly under either improvement.

\emph{The equilibrium comparison.}
Proposition~\ref{prop:existence}'s crossing argument applies in both environments. Since the improved frontier raises $V^*$, it lowers $F=C_T-\phi V^*/T$ pointwise. Its greatest zero is therefore weakly higher. The identity $V=TC_T(T)/\phi$ gives the verified-value comparison.
\end{proof}

\begin{proof}[\textbf{Proof of Proposition~\ref{prop:submissionfees}}]\pagetarget{proof_prop:submissionfees}
Pairwise IC restricts the \emph{difference} between fees, while IR and nonnegative fees restrict their levels. Implementation is possible exactly when these two intervals overlap. We compute the intervals, construct fee levels, and handle the null bundle separately.

For $0<x_L<x_H$, write $s_L=s(x_L)$ and $s_H=s(x_H)$. Pairwise IC gives $\Delta_L\leq s_H-s_L\leq\Delta_H$: the extra high-scope fee deters the weak type but leaves the strong type willing to separate. IR and nonnegativity give $0\leq s_L\leq U_L^L$ and $0\leq s_H\leq U_H^H$, so feasible differences lie in $[-U_L^L,U_H^H]$. The intervals intersect exactly under~\eqref{eq:fee-implementability}. Choose $d$ in the intersection and set $s_L=\max\{-d,0\}$, $s_H=\max\{d,0\}$. For off-menu scope, assign review depth and fee zero; their utility is $-c\leq0$, so the on-menu options remain optimal.
 
\emph{The null-bundle case.}
If $z_L=(0,0)$, the normalization $s(0)=0$ requires $s_L=0$, and $U_L^L=U_H^L=0$. The conditions are then $s_H\geq\max\{0,U_L^H\}$ and $s_H\leq U_H^H$, proving \eqref{eq:fee-null-implementability}. Assumption~\ref{ass:u} gives $u_\theta>0$ on the positive high-type bundle, hence $U_H^H>U_L^H$. Together with $U_H^H\geq0$, this ensures the interval is nonempty.
\end{proof}

\begin{proof}[\textbf{Proof of Proposition~\ref{prop:pricing}}]\pagetarget{proof_prop:pricing}
With a price $p$ per unit of reviewer time, researcher and reviewer optimization jointly maximize a price-adjusted surplus. The key issue is global support: the competitive equilibrium maximizes the same objective globally, and the converse requires a price that globally supports the surplus optimum.

Let $W$ denote resource-adjusted surplus and set $\mathcal J_p=W+p[T-L(\varrho)]$. Researchers maximize verified value less drafting cost and $p$ times load; reviewers maximize $pT-\widetilde C(T)$. Adding their payoffs shows a competitive allocation globally maximizes $\mathcal J_{p^*}$. Market clearing gives $T^*=L(\varrho^*)$, while any capacity-feasible alternative has $T-L(\varrho)\geq0$. Hence
\[
 W(\varrho^*,T^*;\tau)
 =\mathcal J_{p^*}(\varrho^*,T^*;\tau)
 \geq\mathcal J_{p^*}(\varrho,T;\tau)
 \geq W(\varrho,T;\tau).
\]
Thus a competitive allocation attains $S^M(\tau)$. Conversely, the capacity constraint binds at a surplus optimum because $\widetilde C_T>0$, and if the optimum globally maximizes $\mathcal J_{p^*}$ for some $p^*\geq0$, separability makes each researcher's choice and reviewer supply optimal at that price, so it is a competitive equilibrium.
 
As in the proof of Lemma~\ref{lem:fi-frontier}, $c_\tau<0$ at positive scope, so a surplus optimum at $\tau_1$ yields weakly greater surplus at $\tau_2>\tau_1$, strictly if it assigns positive scope to some type.
\end{proof}

\end{appendix}

\bibliography{references}
\bibliographystyle{plainnat}
 
\end{document}